\documentclass[apjl]{emulateapj}
\usepackage{rotate}
\usepackage{amsmath}
\usepackage{natbib}
\bibpunct{(}{)}{;}{a}{}{,}

\slugcomment{To be published on The Astrophysical Journal}

\shorttitle{Synchronous radio/X-ray mode switching of PSR B0943$+$10}

 \shortauthors{S.~Mereghetti~et~ al.}

\def\approxgt{\mathrel{\hbox{\rlap{\lower.55ex \hbox {$\sim$}}
        \kern-.3em \raise.4ex \hbox{$>$}}}}
\def\approxlt{\mathrel{\hbox{\rlap{\lower.55ex \hbox {$\sim$}}
        \kern-.3em \raise.4ex \hbox{$<$}}}}
\def \xmm {\emph{XMM-Newton}}
\def\pdot {\dot P}

\def\ltsima{$\; \buildrel < \over \sim \;$}
\def\lsim{\lower.5ex\hbox{\ltsima}}
\def\gtsima{$\; \buildrel > \over \sim \;$}
\def\gsim{\lower.5ex\hbox{\gtsima}}

\def\psr {PSR B0943+10}
\def\src {PSR B0943+10}

\begin{document}

\title{A deep  campaign to characterize the  synchronous radio/X-ray mode switching of PSR B0943$+$10}

\author{S. Mereghetti\altaffilmark{1}, L. Kuiper\altaffilmark{2}, A. Tiengo\altaffilmark{3,1,4}, J.  Hessels\altaffilmark{5,9}, W. Hermsen\altaffilmark{2,9},   K. Stovall\altaffilmark{6}, A. Possenti\altaffilmark{7}, J. Rankin\altaffilmark{8},   P. Esposito\altaffilmark{9}, R. Turolla\altaffilmark{10,11}, D. Mitra\altaffilmark{8,12,13}, G. Wright\altaffilmark{14}, B. Stappers\altaffilmark{14}, 
A. Horneffer\altaffilmark{15},  S. Oslowski\altaffilmark{15,16,17}, M. Serylak\altaffilmark{18,19}, J.-M. Grie{\ss}meier\altaffilmark{20,19}    }

\altaffiltext{1}{INAF-IASF Milano, via E. Bassini 15, I-20133 Milano, Italy}
\altaffiltext{2}{SRON, Netherlands Institute for Space Research, Sorbonnelaan 2, 3584 CA, Utrecht, The Netherlands}
\altaffiltext{3}{Scuola Universitaria Superiore IUSS Pavia, Piazza della Vittoria 15, I-27100 Pavia, Italy}
\altaffiltext{4}{INFN, Sezione di Pavia, via A. Bassi 6, I-27100 Pavia, Italy}
\altaffiltext{5}{ASTRON, The Netherlands Institute for Radio Astronomy, Postbus 2, 7990 AA Dwingeloo, The Netherlands}
\altaffiltext{6}{Department of Physics and Astronomy, University of New Mexico, Albuquerque, NM, USA}
\altaffiltext{7}{INAF - Osservatorio Astronomico di Cagliari, via della Scienza 5, I-09047 Selargius (CA), Italy }
\altaffiltext{8}{Physics Department, University of Vermont, Burlington, VT 05405, USA}
\altaffiltext{9}{Anton Pannekoek Institute for Astronomy, Univ. of Amsterdam, Science Park 904, 1098 XH, Amsterdam, The Netherlands}
\altaffiltext{10}{Dipartimento di Fisica e Astronomia, Universit\`a di Padova, via F. Marzolo 8, I-35131 Padova, Italy}
\altaffiltext{11}{MSSL-UCL, Holmbury St. Mary, Dorking, Surrey RH5 6NT, UK}
\altaffiltext{12}{National Centre for Radio Astrophysics, Ganeshkhind, Pune 411 007,  India }
\altaffiltext{13}{Janusz Gil Institute of Astronomy, University of Zielona G\'ora, ul. Szafrana 2, 65-516 Zielona G\'ora, Poland}
\altaffiltext{14}{Jodrell Bank Centre for Astrophysics, School of Physics and Astrophysics, University of Manchester,  Manchester, M13 9PL, UK}
\altaffiltext{15}{Max-Planck-Institut f{\"u}r Radioastronomie, Auf dem H{\"u}gel 69, 53121, Bonn, Germany}
\altaffiltext{16}{Fakult{\"a}t f{\"u}r Physik, Universit{\"a}t Bielefeld, Postfach 100131, 33501, Bielefeld, Germany}
\altaffiltext{17}{currently at Centre for Astrophysics and Supercomputing, Swinburne University of Technology, Mail H39, PO Box 218, VIC 3122, Australia}
\altaffiltext{18}{Department of Physics \& Astronomy, University of the Western Cape, Private Bag X17, Bellville 7535, South Africa}
\altaffiltext{19}{Station de Radioastronomie de Nan\c{c}ay, Observatoire de Paris, PSL Research University, CNRS, Univ. d'Orl\'{e}ans, OSUC, 18330 Nan\c{c}ay, France}
\altaffiltext{20}{ LPC2E - Universit\'{e} d'Orl\'{e}ans, CNRS, 45071 Orl\'{e}ans, France}

\begin{abstract}
We report on simultaneous X-ray and radio observations of the mode-switching pulsar \psr\  obtained with the {\it XMM-Newton} satellite and the LOFAR, LWA and Arecibo radio telescopes  in November 2014. We confirm the synchronous X-ray/radio switching between a radio-bright (B) and a radio-quiet (Q) mode, in which the X-ray flux is a factor $\sim2.4$ higher than in the  B-mode. 
We discovered X-ray pulsations, with pulsed fraction of  38$\pm$5\% (0.5-2 keV),  during the  B-mode, and confirm their presence in Q-mode, where the pulsed fraction increases with energy from $\sim$20\%  up to $\sim$65\% at 2 keV. 
We  found marginal evidence for  an  increase in the X-ray pulsed fraction during B-mode on a timescale of hours.
The Q-mode X-ray spectrum  requires a fit with a two-component model (either a power-law plus blackbody or the sum of two blackbodies), while the B-mode spectrum is well fit by a single blackbody (a single power-law  is rejected).  
With a maximum likelihood analysis, we found that in  Q-mode the pulsed emission has  a thermal blackbody spectrum with temperature $\sim3.4\times10^6$ K and the unpulsed emission is a power-law with photon index $\sim$2.5, while during B-mode both the pulsed and unpulsed emission can be fit by either a blackbody or   a power law with similar values of temperature and photon index.  
A {\it Chandra} image shows no evidence for diffuse X-ray emission. 
These results support  a scenario in which both  unpulsed non-thermal emission, likely of magnetospheric origin,   and pulsed thermal emission from a small polar cap ($\sim$1500 m$^2$) with a strong  non-dipolar magnetic field ($\sim10^{14}$ G),  are present during both radio modes and  vary in intensity in a correlated way. This is broadly consistent with the predictions of the partially screened gap model and does not necessarily imply global magnetospheric rearrangements to explain the mode switching. 
   
\end{abstract}


\section{Introduction}
  
Shortly after their discovery, radio pulsars were convincingly interpreted as rapidly rotating neutron stars with a very strong magnetic field, whose rotation and magnetic axes may differ. 
Radio pulsar profiles were early found to be very stable in time, so  it was surprising to find that some were not, even assuming several  stable forms, along with nulling, drifting and other pulse-sequence  phenomena  \citep[e.g.][]{ran86}.
We now know that radio pulsar emission displays a wide range of variations on almost all intensity and time scales, from sparse bursts or nulling, to multi-decade fluctuations. 
Remarkably, some objects exhibit mode changes (or switches): transitions between otherwise stable states with distinct pulse shapes, flux densities, polarization properties, and sometimes different slow-down rates \citep{kra06,lyn10}.
The study of these sources is very important, as they provide glimpses into the dynamics of the neutron star magnetosphere and the poorly understood physics of the pulsar radio emission \citep[e.g.,][]{sob15}.

Here we concentrate on \psr\ which, being the prototypical  mode-switching radio pulsar, has been extensively studied in the radio band and is  a key target to investigate in more details  the high-energy variability of pulsars.  Its   timing parameters  
($P=1.1$ s,   $\pdot$=3.5$\times$10$^{-15}$ s s$^{-1}$) imply, under the usual   assumptions, 
a characteristic age of  $\tau$=$P/(2\pdot$) = 5 Myr, a  dipolar surface magnetic  field $B$=4$\times$10$^{12}$ G, and a  rotational energy loss rate $\dot{E}_{rot}$ = 10$^{32}$ erg s$^{-1}$. 
Detailed modelling of the radio pulse profiles and polarization indicate  that \src\ is a nearly aligned rotator (angle between the rotation and magnetic axis  $\sim$15$^{\circ}$) seen nearly pole-on  \citep{des01}.
Its distance, based on the dispersion measure, and the  Galactic electron density distribution of \citet{cor02},  is $\sim$630 pc.

In the radio band, \src\ alternates between two different states: when it is in the so-called B  (bright) mode, the radio emission displays a regular pattern of drifting subpulses, while it is chaotic, and on average fainter, when in the Q (quiescent) mode \citep{sul84,ran06}.
The phenomenon of drifting subpulses, observed in many radio pulsars, is believed to originate  from a system of sub-beams of radio emission rotating around the magnetic axis \citep{rud75}.
The existence of two modes of emission in \src\ indicates that such a structure is subject to some instability of unclear origin.

Two short \xmm\ observations, carried out in 2003, showed that \src\ is a  faint X-ray source, with a 0.5-8 keV flux of   $\sim5\times10^{-15}$ erg cm$^{-2}$ s$^{-1}$ \citep{zha05}. For a  distance of 630 pc (which we adopt in the whole paper), this corresponds to a luminosity L$_X\sim2\times10^{29}$ erg s$^{-1}$, and implies an X-ray efficiency  in line with that of other rotation-powered pulsars of comparable characteristic age \citep{pos12}. 

A deeper study of the X-ray properties of  \psr\  was performed by \citet{her13}, who used  five \xmm\ pointings, supplemented by simultaneous radio observations with LOFAR and the GMRT. This first multi-wavelength campaign was carried out in 2011 November-December   and provided a useful exposure of  about 100 ks.  Most importantly, using  the mode-change times derived from the radio observations, it was possible to analyze separately the X-ray data of  the Q and B  modes. Quite surprisingly, it was discovered that the X-ray properties in the two modes are  different. 
The X-ray flux is larger by more than a factor two during the Q-mode (when the radio flux is lower by roughly a factor of two at low frequencies; \citet{sul84}). X-ray pulsations at the rotation period of 1.1 s were detected for the first time, but only during the Q-mode. The  Q-mode X-ray spectrum was well fit by the sum of a blackbody and a power law, with single component models clearly rejected. The spectrum of the fainter  B-mode was   less constrained and could be described equally well by  either a power law or a blackbody. From the analysis of the pulsed spectrum in the Q-mode and the timing properties, \citet{her13} concluded that, during the B-mode, \src\ emits only an unpulsed non-thermal component and that the higher flux in Q-mode is caused by the addition of a thermal component with a 100\% pulsed fraction. 

This interpretation challenges the  geometry  of \psr\ derived from the radio observations, which predicts a smaller  modulation of the thermal emission observed from the hot polar cap.  According to \citet{sto14},  a strongly modulated thermal component can be obtained with beamed emission from a magnetic atmosphere or  with an offset dipole geometry. \citet{her13} proposed instead an interpretation based on time-dependent scattering or absorption in the magnetosphere.
This requires some global and rapid rearrangement of the pulsar magnetosphere to explain the different X-ray properties in the two modes. 

A reanalysis of the  2011 \xmm\ observations  was  carried out by \citet{mer13}, who concluded that the data are also consistent with the possibility that a constant, or slightly modulated thermal emission is present in both modes and the flux increase in  the Q-mode is caused by the appearence of a pulsed non-thermal component.
     
To study the remarkable correlated X-ray/radio variability of \psr\ in more detail,  and possibly distinguish between the different interpretations, we  obtained new X-ray  observations in November 2014, within an \xmm\ Large Program with simultaneous radio monitoring provided by the  Low-Frequency Array (LOFAR),    Long Wavelength Array (LWA), and   Arecibo radio telescopes.  
To assess the possible contribution of a pulsar wind nebula to the unpulsed non-thermal X-ray emission seen in \psr , we also obtained the first high spatial resolution X-ray image of this pulsar using  the   {\it Chandra X-ray Observatory}.  In this paper  we concentrate on the results from the new X-ray data.  Further studies of the radio observations, as well as a joint analysis of the whole \xmm\ data set will be presented in  future works. All the errors are at 1 $\sigma$, unless specified differently.

\begin{figure*}
\includegraphics[angle=-90,width=20cm]{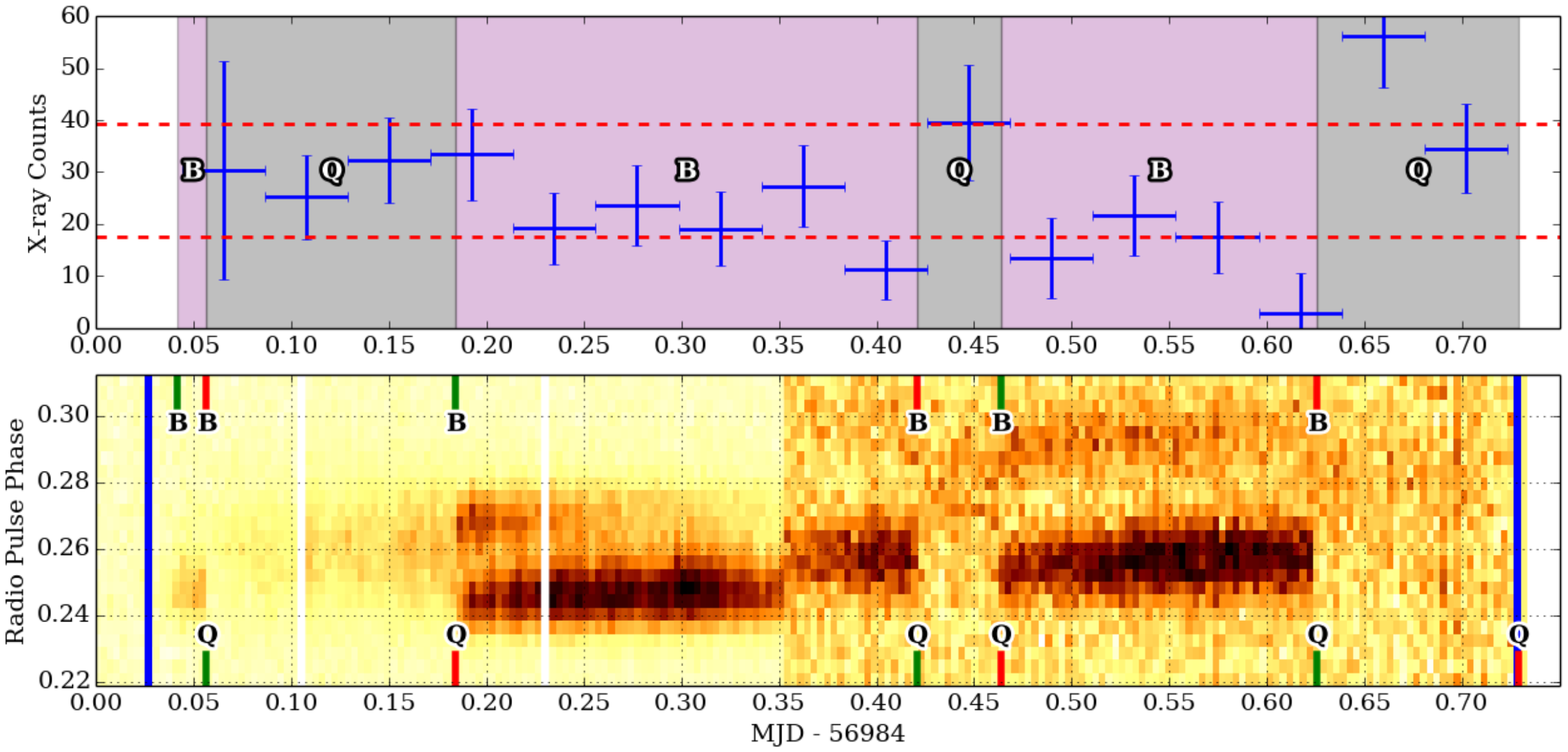}
\vspace{-2.5cm}
\caption{\label{fig_radio_2014} Radio and X-ray lightcurves for Session 5.
{\it Top panel}: 
\xmm\ EPIC pn counts from \psr\ in the 0.2-10 keV band, binned in 3670-s intervals and corrected for the exposure reduction due to the exclusion of high-background periods. 
The radio-identified B/Q-modes are shown by the shaded regions.   
The dashed lines show the average B/Q-mode count levels determined from the entire 2014 data set.
{\it Bottom panel}: 
Radio pulse profiles, each 300\,s, zoomed-in on a narrow range around the main pulse peak.
The sudden transition in profile shape and noise properties at MJD = 56984.35 is when observing coverage shifted from LOFAR (observing at $\sim 150$\,MHz) to LWA (observing at $\sim 60$\,MHz).  
Note that the observed brightness of PSR~B0943+10 is modulated by both intrinsic effects (the mode switching) as well as the effective sensitivity of the telescope due to the changing source elevation.  The start/stop times of B/Q-modes are indicated by the green/red ticks at the top/bottom of the panel.  The vertical blue bars indicate {\it XMM-Newton} observation start/stop times.
}

\end{figure*}

\section{Radio observations and analysis}
\label{sec_radio}

\psr\ is a famously steep-spectrum radio pulsar, with a flux density $S\propto \nu^{-2.6}$ at frequencies \gtsima $80$\,MHz \citep{bkk+15}.
As such, it is best observed at low radio frequencies ($< 1$\,GHz).
This naturally led us to use the LOFAR and LWA telescopes, because of their high sensitivity at low frequencies and also because their complementary geographical locations allowed us to continuously track \psr\ for $\sim 16$\,hr, using also Arecibo to bridge the times when \psr\ was low on the horizon for LOFAR and LWA (the sensivity of both of these aperture array telescopes decreases substantially at low
source elevation).  
In Table~\ref{tab_radiolog}, we present a summary of all the radio observations used to determine the B/Q-mode times
during the \xmm\ observations\footnote{In this table we include only the specifics of LOFAR observations directly used for determining  the modal times.  Redundant data is available from LOFAR stations  UK608, FR606, SE607, and DE601--DE605 -- with roughly the same start  times and durations.}.

\subsection{LOFAR}

\psr\ was observed with the Low-Frequency Array \citep{sha+11,hwg+13}
for a total of $\sim 56$\,hrs.  The LOFAR international stations in
the United Kingdom, France, Sweden and Germany (stations: UK608,
FR606, SE607, and DE601--DE605) were used because these provide
adequate sensitivity for separating \psr's B and Q-modes and are less
over-subscribed than the central part of the LOFAR interferometric
array, which is located in the Netherlands.  Also, having multiple
stations record simultaneously greatly lowers the risk that there will
be a gap in the radio coverage during the  \xmm\ observations.

Each LOFAR international station includes 96 highband antenna (HBA)
tiles, which are sensitive to radio waves in the $110-250$\,MHz range
(though not the whole frequency range at once). The signals from
individual HBA elements are coherently summed at the station,
producing a combined field of view which is effectively like that of a
single-dish radio telescope. The station signals from UK608 
were recorded locally with an {\tt Artemis} system \citep{kar15}. The
signals from SE607  were recorded with the {\tt LuMP}
software\footnote{https://github.com/AHorneffer/lump-lofar-und-mpifr-pulsare}
on local recording machines, while the signals from the German
stations were transmitted via the German LOFAR fibre network to the MPIfR
in Bonn and recorded with {\tt LuMP} there.  For each station, we
recorded 95.3\,MHz of bandwidth (using all 488 `beamlets', of
195.3125-kHz each, in the LOFAR 8-bit station mode), at a center
frequency of 149.9\,MHz.

Initial data analysis and radio frequency interference (RFI) excision
was done using the {\tt PSRCHIVE} suite of pulsar software
\citep{sdo12}, and associated scripts from {\tt
  Coastguard} \footnote{https://github.com/plazar/coast\_guard/}  
 \citep{laz16}.  
 We folded these data sets using an up-to-date rotational ephemeris
derived from ongoing monitoring of \psr\ using the Lovell telescope
(see Table~\ref{tab_ephem}), but using a contemporaneous dispersion
measure derived from the LWA data sets to dedisperse the data.

\subsection{Arecibo}

The Arecibo observations were conducted using the 305-m telescope and
327-MHz receiver system in Puerto Rico.  The pulse sequences were
acquired using four Mock spectrometers covering adjacent 12.5-MHz
portions of the total 50-MHz band.  Arecibo rise-to-set observations
of about 2.4\,hr or, equivalently, 8000 pulses were carried out for
each session apart from that of November 21, which was truncated by
scheduling exigencies.  These observations were processed into
calibrated polarimetric pulse sequences with milli-period time
resolution, wherein the modal transitions could readily be identified
by visual inspection and then confirmed by fluctuation-spectral
analysis (apart from the final session wherein persistent RFI made
identification difficult).

\subsection{Long Wavelength Array}

The Long Wavelength Array (LWA) data were taken using the first station of the LWA (LWA1; \citet{tek+12,ecd+13}), which consists of 256 dipole antennas sensitive to radio waves in the 10--88 MHz range.
LWA1 is co-located with the Very Large Array in New Mexico, USA. 
\src\ was observed using LWA1 for a total of $\sim$72 hrs in raw beam-forming mode using 2 beams (each having 2 tunings) in `split bandwidth' mode with center frequencies of 57.15, 64.5, 71.85, and 79.2 MHz.
The observation setup was similar to other LWA1 pulsar observations described in \citet{sto15}, however, in order to reduce the data size due to the long observing duration, data were recorded
with a reduced sample rate of  9.8M Samples $\mathrm{s^{-1}}$. 
The resulting data were then coherently de-dispersed and divided into 256 frequency channels and folded into 1024 profile bins with 60-s integrations using the {\tt dspsr} software package \citep{sb11}. RFI was excised using a median zapping routine from {\tt  PSRCHIVE }  \citep{sdo12} followed
by manual zapping of RFI. The data were reduced to 16 frequency channels for subsequent analysis.

\subsection{Mode determination}

\psr\ was observed in the first half of each \xmm\ observation by LOFAR and in the second half by LWA.  Arecibo observed near the midpoint of the sessions and provided an important bridge in sensitivity when \psr\ was low on the horizon for both LOFAR and LWA (see Table~\ref{tab_radiolog}).  
We combined the radio data into pulse profile stacks as a function of time, using the {\tt PSRCHIVE} archive
format to store and manipulate the data.  
The individual pulses were summed in 5-min blocks of time to increase the signal to noise ratio
(S/N).  Figure~\ref{fig_radio_2014} shows a representative sample of
the data for session \#5 -- alongside the corresponding \xmm\ X-ray
lightcurve.  The radio profile stacks for all observing sessions are
shown in Appendix A.  Each 5-min integration was qualified as being
either B or Q-mode, based on the S/N and pulse morphology.  We note
that the pulse profile morphology of \psr\ rapidly evolves at low
radio frequencies \citep{bil14}.  Therefore, the LWA ($40-80$\,MHz),
LOFAR ($110-190$\,MHz) and Arecibo ($302-352$\,MHz) profile
morphologies are different. Table~\ref{tab_modes} provides a summary
of all modal time ranges and their durations.  The shortest observed
mode instance lasted 22\,min (B-mode at the beginning of the November
23 observation, but this is a lower limit to the actual duration since
the previous mode transition was not observed). The longest instance
was longer than 17\,hr (Q-mode of the last observation); this is, to our knowledge, the longest mode duration ever reported for this pulsar. 
We have determined the mode instances using a time resolution of 5-min because the S/N of the profiles using shorter integrations is arguably too low to make a robust mode identification.  We note that while some
mode-changing pulsars have in some cases shown mixing between the
modes, making any such identification dubious, this is not the case
for \psr. Given the low X-ray count rate of \psr\ (at most
$\sim0.011$\,ct\,s$^{-1}$), and the fact that the modes typically last
hours, the uncertainty on the times of the mode transitions due to the
adopted 5-min resolution is negligible.

\begin{table}[htdp]
\caption{Radio Observations of \psr\ in 2014}
\label{tab_radiolog}
\begin{center}
\scalebox{0.78}{
\begin{tabular}{lcc}
\hline
\hline
Telescope & Start Time & End Time  \\
 & (UT) & (UT) \\
\hline
\multicolumn{3}{c}{Session 1} \\
\hline
LOFAR/DE601 &  Nov 01 01:24 &  Nov 01 04:18 \\
LOFAR/DE601 &  Nov 01 04:24 &  Nov 01 07:18 \\
LOFAR/DE601 &  Nov 01 07:24 &  Nov 01 10:18 \\
LOFAR/DE605 &  Nov 01 10:24 &  Nov 01 12:38 \\
LWA &  Nov 01 08:32 &  Nov 01 18:30 \\
Arecibo &  Nov 01 10:20 &  Nov 01 12:44 \\
\hline
\multicolumn{3}{c}{Session 2} \\
\hline
LOFAR/DE601 &  Nov 03 00:21 &  Nov 03 03:15 \\
LOFAR/DE601 &  Nov 03 03:21 &  Nov 03 06:15 \\
LOFAR/DE601 &  Nov 03 06:21 &  Nov 03 09:15 \\
LOFAR/DE601 &  Nov 03 09:21 &  Nov 03 12:15 \\
LWA &  Nov 03 09:07 &  Nov 03 19:05 \\
Arecibo &  Nov 03 10:14 &  Nov 03 12:32 \\
\hline
\multicolumn{3}{c}{Session 3} \\
\hline
LOFAR/DE601 &  Nov 05 00:13 &  Nov 05 03:07 \\
LOFAR/DE601 &  Nov 05 03:13 &  Nov 05 06:07 \\
LOFAR/DE601 &  Nov 05 06:13 &  Nov 05 09:07 \\
LOFAR/DE601 &  Nov 05 09:13 &  Nov 05 12:07 \\
LWA &  Nov 05 09:02 &  Nov 05 19:50 \\
Arecibo &  Nov 05 10:05 &  Nov 05 12:29 \\
\hline
\multicolumn{3}{c}{Session 4} \\
\hline
LOFAR/DE601 &  Nov 20 23:37 &  Nov 21 02:31 \\
LOFAR/DE601 &  Nov 21 02:37 &  Nov 21 05:31 \\
LOFAR/DE601 &  Nov 21 05:37 &  Nov 21 08:31 \\
LOFAR/DE601 &  Nov 21 08:37 &  Nov 21 11:31 \\
LWA &  Nov 21 07:47 &  Nov 21 17:45 \\
Arecibo &  Nov 21 09:02 &  Nov 21 10:00 \\
\hline
\multicolumn{3}{c}{Session 5} \\
\hline
LOFAR/DE601 &  Nov 22 23:34 &  Nov 23 02:28 \\
LOFAR/DE601 &  Nov 23 02:34 &  Nov 23 05:28 \\
LOFAR/DE601 &  Nov 23 05:34 &  Nov 23 08:28 \\
LOFAR/DE601 &  Nov 23 08:34 &  Nov 23 11:28 \\
LWA &  Nov 23 07:37 &  Nov 23 17:35 \\
Arecibo &  Nov 23 08:55 &  Nov 23 11:15 \\
\hline
\multicolumn{3}{c}{Session 6} \\
\hline
LOFAR/DE601 &  Nov 24 23:32 &  Nov 25 02:26 \\
LOFAR/DE601 &  Nov 25 02:32 &  Nov 25 05:26 \\
LOFAR/DE601 &  Nov 25 05:32 &  Nov 25 08:26 \\
LOFAR/DE601 &  Nov 25 08:32 &  Nov 25 11:26 \\
LWA &  Nov 25 08:02 &  Nov 25 18:30 \\
Arecibo &  Nov 25 08:45 &  Nov 25 11:12 \\
\hline
\multicolumn{3}{c}{Session 7} \\
\hline
LOFAR/FR606 &  Nov 26 23:39 &  Nov 27 10:49 \\
LOFAR/DE601 &  Nov 27 07:37 &  Nov 27 10:31 \\
LOFAR/DE601 &  Nov 27 10:37 &  Nov 27 13:31 \\
LWA &  Nov 27 08:02 &  Nov 27 18:30 \\
Arecibo &  Nov 27 08:39 &  Nov 27 11:03 \\
\hline
\hline
\end{tabular}
}
\end{center}
\label{default}
\end{table}

\begin{table}[htbp!]
\caption{Q- and B-mode time intervals in    November 2014
\label{tab_modes}}
\begin{center}
\scalebox{0.85}{
\begin{tabular}{ccrrcc}
\hline 
   \multicolumn{2}{c}{Start (UT)}      &   \multicolumn{2}{c}{End (UT)}                 & Duration  & Mode \\
     Day &HH:MM       &   Day &HH:MM            &      hr      &   \\
\hline
\multicolumn{6}{c}{Session 1} \\
\hline
  1 & 02:55 &  1 & 08:56 &   6.02      & B  \\
  1 & 08:56 &  1 &13:58 &   5.03      & Q  \\
  1 & 13:58 & 1 &18:28 &   4.51      & B  \\
\hline
\multicolumn{6}{c}{Session 2} \\
\hline
  3 & 02:34 &  3 & 06:30 &   3.93      & B  \\
  3 & 06:30 &  3 & 07:23 &   0.87      & Q  \\
  3 & 07:23 &  3& 14:04 &   6.69      & B  \\
  3& 14:04 &  3 &18:57 &   4.89      & Q  \\
\hline
\multicolumn{6}{c}{Session 3} \\
\hline
  5  &02:32 &  5 & 05:57 &   3.43      & Q  \\
  5 & 05:57 &  5& 12:12 &   6.23      & B  \\
  5& 12:12 &  5 &13:10 &   0.97      & Q  \\
  5& 13:10 &  5 &17:32 &   4.37      & B  \\
  5& 17:32 &  5 &19:45 &   2.22      & Q  \\
\hline
\multicolumn{6}{c}{Session 4} \\
\hline
 21 & 01:36 & 21 & 08:23&   6.79      & B  \\
 21 & 08:23 & 21 &11:39 &   3.26      & Q  \\
 21 &11:39 & 21 &17:38 &   5.98      & B  \\
\hline
\multicolumn{6}{c}{Session 5} \\
\hline
 23 & 00:59 & 23 & 01:21 &   0.36      & B  \\
 23 & 01:21 & 23 & 04:25 &   3.06      & Q  \\
 23 & 04:25 & 23 &10:06 &   5.68      & B  \\
 23& 10:06 & 23 &11:08 &   1.03      & Q  \\
 23 &11:08 & 23 &15:01 &   3.89      & B  \\
 23 &15:01 & 23 &17:29 &   2.47      & Q  \\
\hline
\multicolumn{6}{c}{Session 6} \\
\hline
 25  &00:51 & 25  &02:14 &   1.38      & B  \\
 25 & 02:14 & 25  &05:14 &   3.00      & Q  \\
 25  &05:14 & 25 &13:21 &   8.13      & B  \\
 25 &13:21 & 25 &15:46 &   2.41      & Q  \\
 25 &15:46 & 25& 18:21 &   2.59      & B  \\
\hline
\multicolumn{6}{c}{Session 7} \\
\hline
 26& 23:41 & 27& 00:15 &   0.56      & B  \\
 27 & 00:15 & 27 &17:33 &  17.30      & Q  \\
\hline
\end{tabular}
}
\end{center}
Note: The start/stop times have an uncertainty of 5 min. They refer to  the  intervals in which we observed \psr\ in a given mode and do  not necessarily correspond to the actual start/stop times of complete mode instances.  
\end{table}

\section{X-ray  observations and analysis}

A log of the  2014 X-ray observations of \psr\   is given in Table \ref{tab_obs_xmm}. 
Seven whole satellite orbits were devoted to this campaign, corresponding to a total   of  $\sim$120 hr,  to be compared to the six observations of 6 hr each of the  2011 campaign.
Due to the faintness of the source, only the data obtained with the EPIC instrument can be used. 
EPIC consists of one camera based on pn CCDs and two cameras based on MOS CCDs, covering  the 0.2-12 keV energy range \citep{str01,tur01}. During all  observations, the pn camera was operated in full frame  mode, which provides a time resolution of 73 ms. For the two MOS cameras, we used the small window mode, which is the MOS imaging mode with the highest time resolution (0.3 s).  For the three cameras we used the thin optical filter. These settings of the EPIC instruments are the same that were used in the 2011   campaign.  

\begin{table}[htbp!]
\caption{\xmm\ observations of \psr\ in 2014
\label{tab_obs_xmm}
}
\begin{center}
\begin{tabular}{cccc}
\hline 
Session &   Obs. ID                   &  Start     &  End             \\
             &      &    (UT)  pn/MOS &  (UT)      pn/MOS      \\
\hline
 1&  0743950101    &  Oct 31  23:37/23:14 &  Nov  01  18:14/18:18                   \\
 2&  0743950201    &  Nov  03 02:39/02:06 &  Nov   03 18:52/18:56                  \\
 3&  0743950301    &  Nov  05 02:54/03:58 &  Nov     05 19:35/19:39                \\
 4&  0743950401    &  Nov  21  02:32/02:09 &  Nov   21 17:33/17:37                 \\
 5&  0743950501    &  Nov   23 01:01/00:38 &  Nov    23 17:25/17:29               \\
 6&  0743950601    &  Nov 25 01:37/01:15&  Nov   25 18:18/18:22                   \\
 7&  0743950701    &  Nov  27 03:44/02:11&  Nov   27 19:40/19:40                   \\  
 \hline
\end{tabular}
\end{center}
\end{table}

All the observations were affected, in different measure, by periods of high particle background. These were excluded by adopting different filtering criteria, depending on the type of analysis, as described  below. 
Nevertheless, the 2014  data  provide a significant increase in the number of counts collected from \psr\ with respect to the 2011 observations (a factor $\gsim$2 for the Q-mode and $\gsim$3 for the B-mode).

\subsection{Timing}
\label{sec_timing}

The pn and MOS counts for the timing analysis were  extracted from a circular region with radius of 15$''$  and the times of arrival were converted to the solar system barycenter.  We excluded the parts of the observations  affected by a high background level by selecting only time intervals in which the EPIC pn count rate in the 10-12 keV range was below 1.2 counts s$^{-1}$. This resulted in exposure times of  119 ks and  175 ks, for the Q and B-mode, respectively. 

\begin{figure*}
\includegraphics[angle=90,width=19cm]{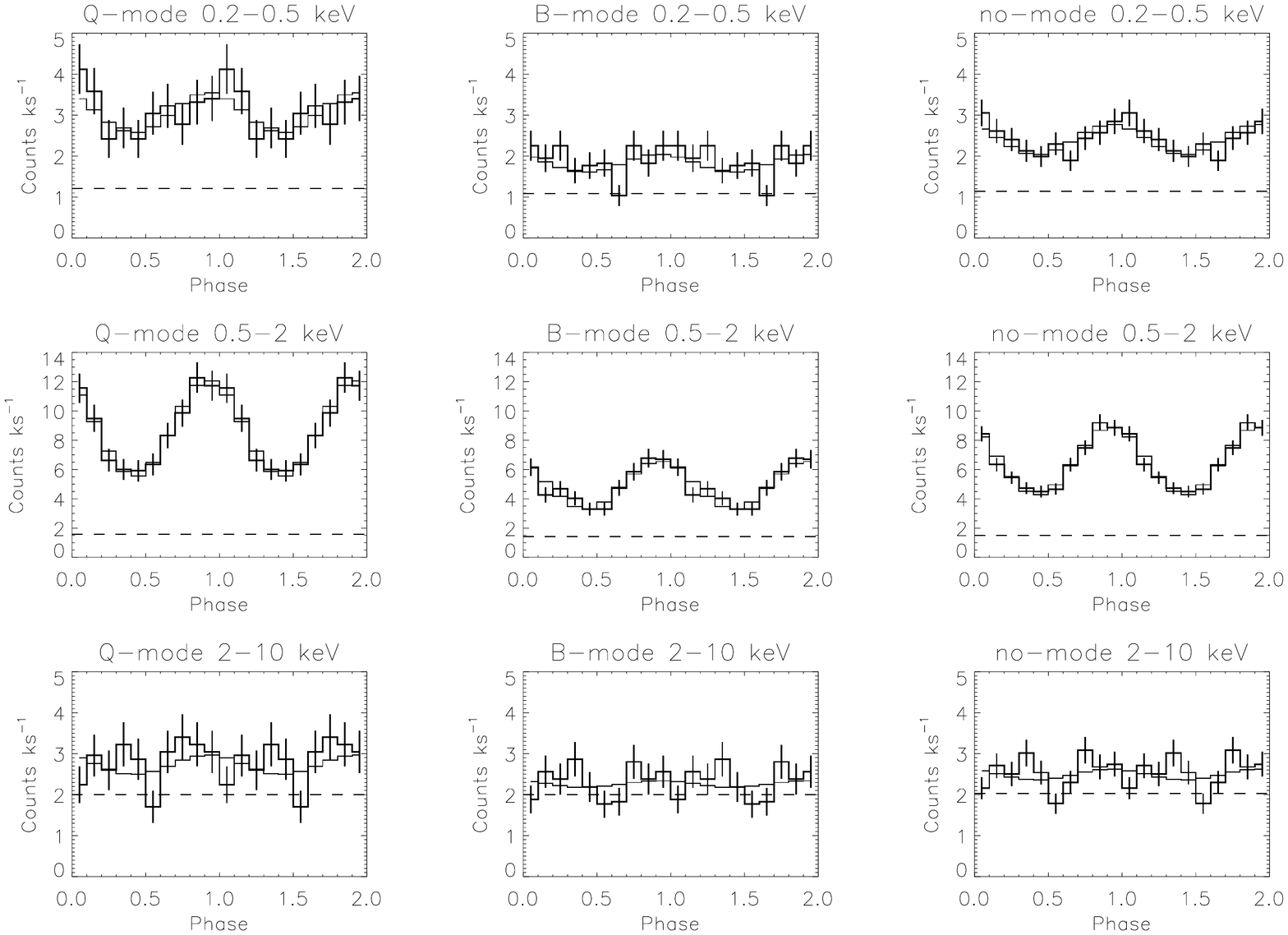}
\caption{\label{fig_folded_2014} Folded pn+MOS light curves. The thin solid lines are the best-fit sinusoidal functions (with phase fixed to that determined from the Q-mode 0.5-2 keV pulse profile). The horizontal dashed lines indicate the background level.}
\end{figure*}

The pulse phases of the \psr\ counts were computed using  the same ephemeris  adopted for the radio data (Table~\ref{tab_ephem}).
In Fig.~\ref{fig_folded_2014} we show the folded pulse profiles in different energy ranges for the Q and B-mode time intervals, as well as those corresponding to the whole observation. The latter differs from the sum of the B and Q time intervals, since it includes also   $\sim$17 ks for which information on the radio mode is missing. 

The pulsations are clearly detected     in the 0.5-2 keV energy range during both radio modes, with statistical significance, computed using the Rayleigh test statistics $Z^2_1$ \citep{buc83},  of 8.4$\sigma$ in the Q-mode and 6.3$\sigma$ in the B-mode. The 0.5-2 keV  pulse profiles  of the two modes are similar. They are broadly sinusoidal, with  pulsed fractions, defined as the amplitude of the sinusoid divided by the average value, of   45$\pm$6\%  (Q-mode) and 47$\pm$6\%  (B-mode), and aligned in phase. These pulsed fractions are consistent with the more precise values obtained in Sect.~\ref{sec_ml} with a Maximum Likelihood analysis.   By a cross-correlation of the two light curves, we could set a 90\% c.l. upper limit of 0.08 cycles  on the relative phase of the peak in the B and Q-mode.
No pulsations are  seen above 2 keV, while there is a hint that they could be present also below 0.5 keV. In fact,  although the statistical significance of the modulation in the  0.2--0.5 keV range is only $\sim2\sigma$ in the Q and B mode separately, a level of 3.4$\sigma$ is reached when one considers the whole data set.

\begin{table}[htbp!]
\caption{Ephemeris of \psr\ derived from Lovell telescope monitoring observations
\label{tab_ephem}
}
\begin{center}
\begin{tabular}{lc}
\hline
 Right Ascension (J2000)         &     09$^h$ 46$^m$ 07.786$^s$             \\
 Declination (J2000)              &               09$^{\circ}$ 52$'$ 00.76$''$         \\
$\nu$       (Hz)                       &      0.9109890963(2)     \\
$\dot \nu$ (Hz s$^{-1}$)    &   $-$2.9470(7)$\times10^{-15}$  \\
$\ddot \nu$  (Hz s$^{-2}$)          &   $-$1.38$\times10^{-25}$      \\
Epoch (MJD)                      &   55962.057374  \\ 
Validity range (MJD)       &    54861.014-- 57011.249 \\ 
\hline
\end{tabular}
\end{center}
The coordinates refer to  the X-ray position of \psr , which has an accuracy  of 1$''$ \citep{her13}. Numbers in parenthesis are the 1$\sigma$ errors on the least significant digits.
\end{table}

\subsection{Spectral analysis}
\label{sec_spectral_bis}

As a first step for the  analysis of the phase-averaged spectra of \src , we reprocessed the pn data using the SAS task  {\it epreject}  to reduce the detector noise at the  lowest energies. We then  applied a tight screening  to eliminate periods of high background due to flares of soft protons. For the  extraction of the source spectra we used   circular regions of radius 15$''$, while for the background we used circles of   radius 35$''$ for the pn and  25$''$ for the MOS.   We used   single-pixel events for the pn (PATTERN=0) and single- and multiple-pixel events for the   MOS   (PATTERN $\leq$12). We first combined   the  spectra of the two MOS   cameras  and then summed the spectra of the Q- and B-mode time intervals of all the observations. 
This resulted in four spectra with net live times  in the Q-mode of 67.7 ks (pn) and 113.5 ks  (MOS), and in the B-mode of  90.9 ks (pn) and 162.6 ks (MOS).

For each mode, we simultaneously fitted the pn and MOS spectra in the energy range  0.2-10 keV using XSPEC. 
Following previous works \citep{zha05,her13} and to facilitate comparisons with their results, we fixed the absorbing
column density to $N_{\rm{H}}=4.3\times10^{20}$~cm$^{-2}$. This value corresponds to the dipersion measure of \psr\  assuming a 10\% ionization of the interstellar medium (e.g. \citealt{he13}).

The analysis of the Q-mode spectra showed that  fits with a single power-law or a single blackbody are not acceptable.
With fixed $N_H$, these models gave values of $\chi_\nu^2>2.5$ (for 20 degrees of freedom (dof)), corresponding to  null hypothesis probabilities (nhp) smaller than $2\times10^{-4}$.
Letting $N_H$ free to vary, the single blackbody fit was still rejected (nhp = $2\times10^{-3}$), while a formally acceptable fit was obtained with a power-law ($\chi_\nu^2$/dof=1.4/19, nhp=0.12). However, the latter resulted in a  large power law photon index ($\Gamma=3.3_{-0.2}^{+0.3}$) and an absorption value ($N_H=(1.8\pm0.4)\times10^{21}$ cm$^{-2}$)  much higher than  that expected for \psr\ (the total Galactic column density in this direction is $2.5\times10^{20}$~cm$^{-2}$, \citet{kal05}). 
A good fit to the Q-mode spectra was instead obtained using either a blackbody plus power-law model or the sum of two blackbodies, resulting in the parameters  given in Table~\ref{tab_sp1_bis}.

The B-mode spectra were well fit by  a  blackbody with temperature kT=0.23 keV, while a power-law was clearly rejected ($\chi_\nu^2$/dof=2.1/20,  nhp=3$\times10^{-3}$).
A single power-law gave a statistically acceptable fit (nhp=0.13) only with rather implausible parameters ($\Gamma=4.1_{-0.6}^{+0.8}$,  $N_H=(3.2\pm0.1)\times10^{21}$ cm$^{-2}$). 
Good fits could also be obtained with two-component models, but, contrary to the case of the Q-mode,  the addition of a further component to the single blackbody is not statistically required.

Finally, to assess whether there is a significant spectral difference between the two modes,  we fitted simultaneously the B and Q spectra with a blackbody plus power-law model keeping $\Gamma$, kT, and relative normalization of the two components tied to common values.
In this way we obtained a good fit with the parameters given in the last column of  Table~\ref{tab_sp1_bis} and a scale factor between the B and Q spectra of $f_{B/Q}$=0.42$\pm$0.03.

\begin{table*}
\centering \caption{Results for the total spectra of Q and B modes with standard analysis}
\label{tab_sp1_bis}
\begin{tabular}{lcccccc}
\hline
                                    & Q-mode             & Q-mode            & B-mode            & B-mode                 & Joint fit of B and Q modes \\
                                    &  BB + PL             &         BB + BB    &                BB        & BB + PL                         & BB + PL \\
\hline\\
Photon index            & 2.7$_{-0.2}^{+0.3}$       &      --                         &                  --               &   2.5$\pm$0.4                     &                 2.6$_{-0.2}^{+0.3}$   \\
K$_{PL}^a$           & 2.4$\pm$0.6               &            --                     &                     --            &  0.6$\pm$0.5                &   2.1$\pm$0.5$^e$ / 0.9$\pm$0.2$^f$    \\
F$_{PL}^d$                  &   5.5$\pm$1.1        &        --                               &        --                     & 1.4$\pm$0.4       &  --  \\

kT$_1$ (keV)            &   0.29$\pm$0.03          & 0.14$\pm$0.02        &  0.23$\pm$0.01            &  0.24$\pm$0.03           &   0.27$\pm$0.02  \\
$\alpha_{BB1}^b$    & 1.15$_{-0.40}^{+0.75}$  &  29$_{-10}^{+18}$  &2.2$_{-0.4}^{+0.6}$        &  1.4$_{-0.5}^{+1.1}$  &      1.6$\pm$0.6$^e$ / 0.7$\pm$0.4$^f$  \\
R$_{BB1}^c$  (m)     & 21$_{-4}^{+6}$              &  106$_{-20}^{+29}$  &29$_{-3}^{+4}$             &  23$_{-5}^{+8}$            &      25$\pm$5$^e$ / 16$\pm$4$^f$  \\
F$_{BB1}^d$               &6.4$\pm$1.2          &     4.8$\pm$1.3                   &        --                    &  3.8$_{-0.8}^{+2.4}$     &  -- \\

kT$_2$ (keV)           &   --                              & 0.37$\pm$0.04           &     --                            &    --  &   -- \\
$\alpha_{BB2}^b$     &  --                              & 0.51$_{-0.20}^{+0.33}$   &     --                        &    --       &    --   \\
R$_{BB2}^c$  (m)      &  --                         & 14$_{-3}^{+4}$                     &     --                       &    --       &    --   \\
F$_{BB2}^d$              &  --                           &     7.5$\pm$1.3        &        --                   &  --     & --   \\

F$_{TOT}^d$                &      11.8$\pm$0.5   &   12.2$\pm$0.5                & 5.2$\pm$0.3             &     5.2$\pm$0.3                  & -- \\
$\chi_{\nu}^2$/dof  &   1.24/18              &  0.98/18                             &    1.12/20                 &  1.03/18                 &     1.11/39 \\
N.h.p.                         &    0.22                   &   0.47                                &  0.32                          &    0.42                     &        0.29  \\
  \hline
\end{tabular}
\begin{list}{}{}
\item Joint fits of pn + MOS spectra with fixed $N_H=4.3\times10^{20}$ cm$^{-2}$.  Errors at 1$\sigma$.  
\item[$^{a}$]
 Normalization of the power-law at 1 keV in units of 10$^{-6}$ ph cm$^{-2}$ s$^{-1}$ keV$^{-1}$.
 \item[$^{b}$]
 Blackbody normalization in units of   10$^{-4}$ ph cm$^{-2}$ s$^{-1}$ keV$^{-3}$.
 \item[$^{c}$]
 Blackbody radius for d=630 pc.
 \item[$^{d}$]
Flux  in the 0.5-2 keV range corrected for the absorption, in units of 10$^{-15}$ erg cm$^{-2}$ s$^{-1}$.
\item[$^{e}$]
 Q-mode.
\item[$^{f}$]
 B-mode.

\end{list}
\end{table*}

\begin{figure} 
\vspace{-2cm}
\includegraphics[angle=0,width=9cm]{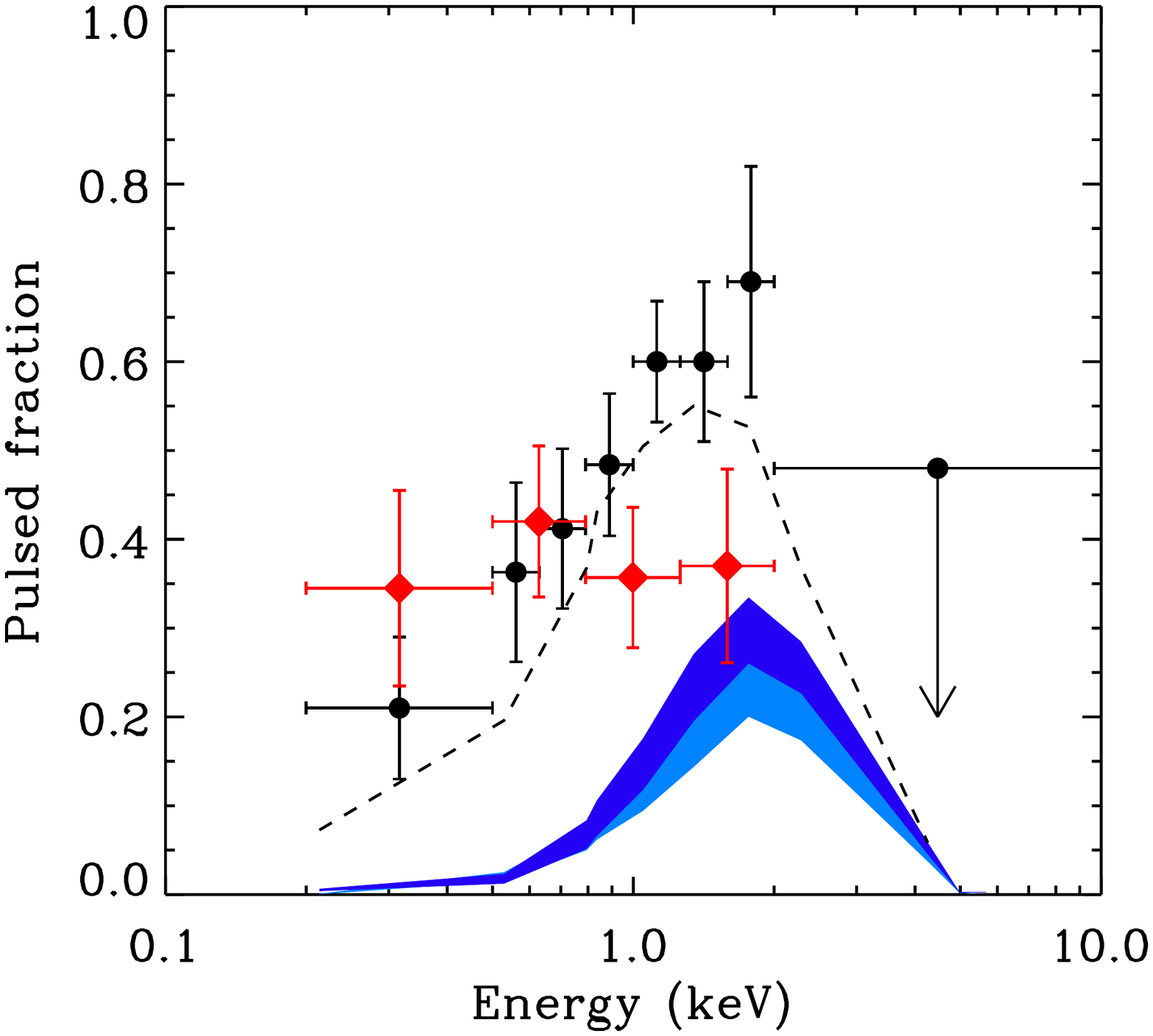}
\vspace{-3cm}
\caption{\label{fig_pf} 
Pulsed fraction as a function of energy for the Q-mode (black dots) and B-mode (red diamonds)  as derived from the 3D-ML analysis. Errors bars are at 1$\sigma$, the upper limit at 2$\sigma$. The blue regions indicate the comparison with model atmosphere predictions for the most likely geometry of \psr\ (see Sect.~\ref{sec_pf}). 
The dashed line is the pulsed fraction computed by  \citet{sto14}, corrected to account for the unpulsed non-thermal emission.
}
\end{figure}


\begin{table*}[htbp!]
\centering \caption{Results for the total spectra of Q and B modes with 2D-ML analysis}
\label{tab_ML2D}
\begin{tabular}{lccccc}
\hline
                       & Q-mode                    & Q-mode                   & B-mode                    & B-mode                  &  Joint fit of B and Q modes \\
                       & BB + PL                   & BB + BB                  & BB                        & BB + PL                 &      BB + PL          \\
\hline\\
Photon index            & 2.4$\pm$0.2                      & --                       & --                        & 2.3$_{-0.5}^{+0.4}$        &  2.4$\pm$0.1 \\
K$_{PL}^a$             & 3.2$\pm$0.7                     & --                       & --                        & 0.8$\pm$0.5                       &   2.8$\pm$0.2$^e$ / 1.2$\pm$0.2$^f$\\
F$_{PL}^d$                 & 7.4$\pm$0.7                     & --                       & --                        & 1.6$\pm$0.4               &        --      \\

kT$_1$ (keV)           & 0.27$\pm$0.04                & 0.11$\pm$0.02   & 0.24$\pm$0.01  & 0.24$\pm$0.03 & 0.26$\pm$0.01 \\

$\alpha_{BB1}^b$        & 1.12$_{-0.49}^{+0.85}$    & 61$_{-32}^{+77}$         & 2.3$_{-0.4}^{+0.5}$    & 1.5$_{-0.6}^{+1.0}$   & 1.7$\pm$0.2$^e$ / 0.8$\pm$0.1$^f$\\
R$_{BB1}^c$ (m)      & 21$_{-5}^{+8}$            & 153$_{-40}^{+97}$        & 29$_{-3}^{+3}$            & 24$_{-5}^{+8}$            &  25$\pm$2$^e$ / 17$\pm$1$^f$ \\

F$_{BB1}^d$          & 4.9$\pm$0.9               & 3.0$\pm$0.4              &   --          & 4.1$\pm$0.6   &  -- \\

kT$_2$ (keV)           & --                        & 0.34$\pm$0.03        & --                        & --                     & --   \\
$\alpha_{BB2}^b$        & --                        & 0.90$_{-0.26}^{+0.36}$   & --                        & --                      & --  \\
R$_{BB2}^c$ (m)      & --                        & 19$_{-3}^{+4}$           & --                        & --                      &   -- \\

F$_{BB2}^d$   & --                        & 9.6$\pm$0.9   & --                        & --              &         --  \\

F$_{TOT}^d$                &      12.3$\pm$1.1  &   12.6$\pm$1.0         & 5.7$\pm$0.3         &     5.7$\pm$0.7                  & -- \\
$\chi_{\nu}^2$/dof     & 0.63/30                   & 1.02/30                  & 1.15/24                   & 0.85/22     &     0.71/56          \\
N.h.p.                 & 0.94                        & 0.44               & 0.28                        & 0.66                     &     0.95 \\
\hline
\end{tabular}
\begin{list}{}{}
\item Joint fits of pn + MOS spectra with fixed $N_H=4.3\times10^{20}$ cm$^{-2}$.  Errors at 1$\sigma$.  
\item[$^{a}$]
 Normalization of the power-law at 1 keV in units of 10$^{-6}$ ph cm$^{-2}$ s$^{-1}$ keV$^{-1}$.
 \item[$^{b}$]
 Blackbody normalization in units of   10$^{-4}$ ph cm$^{-2}$ s$^{-1}$ keV$^{-3}$.
 \item[$^{c}$]
 Blackbody radius for d=630 pc.
 \item[$^{d}$]
Flux  in the 0.5-2 keV range corrected for the absorption, in units of 10$^{-15}$ erg cm$^{-2}$ s$^{-1}$.
\item[$^{e}$]
 Q-mode.
\item[$^{f}$]
 B-mode.
\end{list}
\end{table*}



\begin{table*}[htbp!]
\centering \caption{Results for the pulsed and unpulsed spectra of Q and B modes with 3D-ML analysis}
\label{tab_ML3D}
\begin{tabular}{lcccccc}
\hline
                     & Q-mode  & Q-mode         & B-mode  & B-mode & B-mode  &  B-mode  \\
                     &  pulsed    &  unpulsed      & pulsed          & pulsed               &   unpulsed          &unpulsed  \\
                        &  BB        &  PL               &         BB        & PL               &    BB            &  PL   \\
      \hline\\
 
Photon index      &   -- & 2.5$\pm$0.2       &--  &  2.3$\pm$0.3   & --& 2.4$\pm$0.2 \\
 
K$_{PL}^a$       & --    &  2.9$\pm$0.3            & --  & 0.9$\pm$0.1  & --& 1.4$\pm$0.2 \\
 
kT (keV)         &   0.29$\pm$0.02              & -- & 0.22$_{-0.03}^{+0.04}$ &-- & 0.22$_{-0.02}^{+0.03}$ & --\\
$\alpha_{BB}^b$  & 1.11$_{-0.28}^{+0.39}$    & -- & 1.24$_{-0.64}^{+1.24}$  &-- & 1.88$_{-0.72}^{+1.13}$ & --\\
R$_{BB}^c$  (m) & 20.6$\pm$3.1  & -- & 21.8$\pm$8.3& --& 26.9$\pm$6.6 &--\\
 
F$^d$  & 6.1$_{-0.5}^{+0.6}$ &  6.5$_{-0.5}^{+0.6}$ &2.1$_{-0.3}^{+0.4}$ & 2.0$\pm$0.3     & 3.6$\pm$0.4  & 3.2$\pm$0.4 \\
$\chi_{\nu}^2$/dof  &   1.07/8   & 0.85/8 & 0.57/5 & 0.37/5 & 0.56/5  & 0.86/5\\
N.h.p.                         & 0.38      & 0.56     &  0.72   & 0.87     & 0.73     & 0.51  \\
  \hline
\end{tabular}
\begin{list}{}{}
\item Joint fits of pn + MOS spectra with fixed $N_H=4.3\times10^{20}$ cm$^{-2}$.  Errors at 1$\sigma$.  
\item[$^{a}$]
 Normalization of the power-law at 1 keV in units of 10$^{-6}$ ph cm$^{-2}$ s$^{-1}$ keV$^{-1}$.
 \item[$^{b}$]
 Blackbody normalization in units of   10$^{-4}$ ph cm$^{-2}$ s$^{-1}$ keV$^{-3}$.
 \item[$^{c}$]
 Blackbody radius for d=630 pc.
 \item[$^{d}$]
Flux  in the 0.5-2 keV range corrected for the absorption, in units of 10$^{-15}$ erg cm$^{-2}$ s$^{-1}$.
\end{list}
\end{table*}

\subsection{Maximum likelihood  spectral and timing analysis}
\label{sec_ml}

In this subsection we follow an alternative approach, based on a maximum likelihood (ML) analysis,   which is particularly powerful in the case of faint sources.  This method (2D-ML, in the following) has been used by \citet{her13}  to obtain the source count rates and spectra  in Q- and B-mode for the first radio/X-ray campaign on \psr\ performed in 2011. Here we extend it  to derive also the spectra of the pulsed and unpulsed components of both radio modes (3D-ML). 

Contrary to the traditional analysis, in which the background is estimated from a ``source-free'' region of the image, 
in the 2D-ML   the instrumental point spread function (PSF)  and  a background distribution are simultaneously fitted to the data in the region  including the source of interest, taking into account the Poissonian nature of the counting process. 
At first, the selected events are sorted according to their spatial coordinates $(x,y)$  to produce counts skymaps\footnote{Each event is characterized by its (barycentered) time $t$, spatial coordinates $x$, $y$, energy $E$, pattern $\xi$ and flag $\digamma$. In all the following analysis we used two dimensional pixels of  2\arcsec $\times$ 2\arcsec\ in size, $\xi=[0,4]$ and $\digamma=0$ for both pn and MOS data.}.
The likelihood function $L$ is defined as \\

$L=\ln\left( \prod_{i,j}\ (\mu_{ij}^{N_{ij}} \exp(-\mu_{ij}) /N_{ij}!)\right) =$ \\

~~~~$= \sum_{i,j} N_{ij}\ln(\mu_{ij}) - \mu_{ij} -\ln(N_{ij}!) $,


\medskip
\noindent
where $\mu_{ij}=\beta+\sigma \cdot {PSF}_{ij}$ is the expectation value for pixel $(i,j)$ and $N_{ij}$ is the number of counts measured in pixel $(i,j)$. 
The quantity $L$ is maximized simultaneously with respect to the background parameter $\beta$ and the source scale parameter $\sigma$. 
Since the  PSF is normalized to unity, $\sigma$ gives directly  the total number of net (background free) source counts. 
The second derivative matrix of $L$, evaluated at the maximum value, contains information on the uncertainties of the derived parameters, $\beta$ and $\sigma$.
Applying this optimization scheme for a grid of energy windows 
results in net source counts per energy interval, 
which can be converted to source fluxes by a forward folding fitting procedure assuming a spectral model and using appropriate response  ({\tt arf} and {\tt rmf} files)  and livetime information (dead-time corrected exposure). 

We applied this method, separately for the pn and for the sum of the two MOS,  to extract the time-integrated Q- and B-mode spectra in the  0.2-10 keV  range. Time intervals with high background were excluded using the same cuts of section \ref{sec_timing}. 
For the pn and MOS PSF we used axially symmetric profiles as derived from in-flight calibrations\footnote{ http://www.cosmos.esa.int/web/xmm-newton/calibration-documentation.},   with  parameters appropriate for an energy of 1.5 keV. We modeled the background with a uniform  spatial distribution. As in the previous subsection, we fixed N$_{\hbox{\scriptsize H}}=4.3\times 10^{20}$ cm$^{-2}$.

For the Q-mode, single component power-law or blackbody  models yielded unacceptable $\chi^2$  values, while for the B-mode a single power-law is excluded,  contrary to the 2011 data, which could be described equally well by either  a power-law or a blackbody. A single blackbody, however, adequately describes the data, as does the combination of a blackbody plus either a power-law  or another blackbody. 
These results, summarized in Table~\ref{tab_ML2D}, are fully consistent with those described in the previous subsection.

\begin{figure*}
\begin{center}
\includegraphics[width=9cm,height=18cm,angle=90,bb=120 160 465 760,clip=]{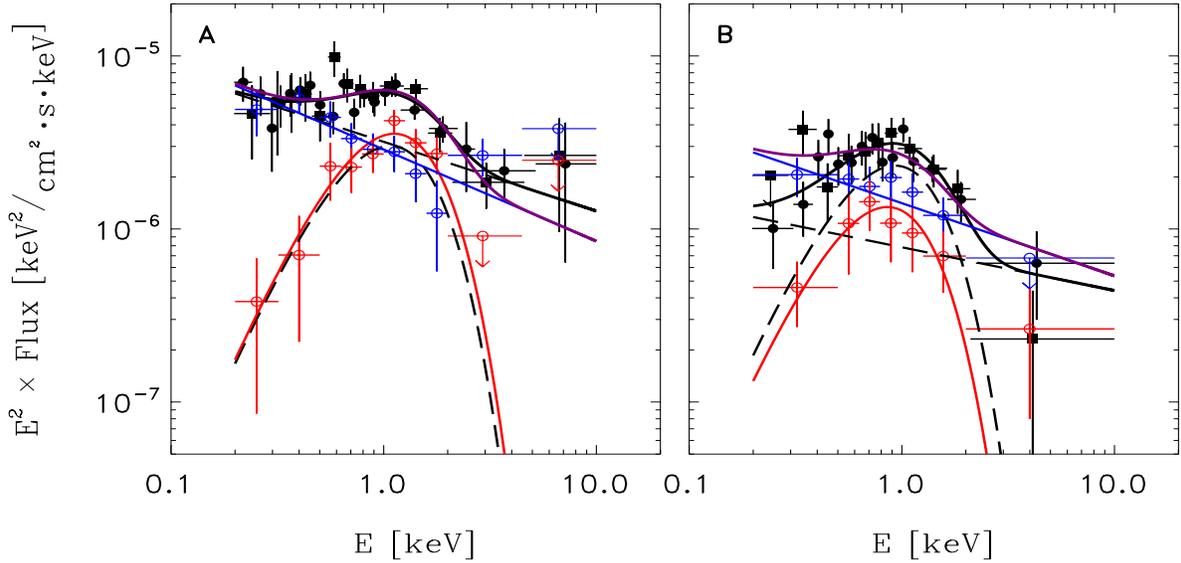}
\caption{\label{fig_sp} Results of the ML spectroscopy for the Q-mode (left panel) and B-mode (right panel). 
Black lines and points refer to the 2D-ML   (circles pn, squares sum of the two MOS, solid lines are the best-fit blackbody plus power law models, dashed lines the individual components).  The results of the 3D-ML analysis are in blue for the unpulsed component (data points and best-fit with a single power-law)  and in red for the pulsed component (data points and best-fit with a single blackbody). The purple line is the sum of the blue and red lines. }
\end{center}
\end{figure*}

Fig. \ref{fig_sp}  shows with black symbols (circles for pn, squares for the summed MOS) the unabsorbed Q- (left) and B-mode (right) spectra for
the best-fit blackbody plus power-law model. The separate spectral components are shown as black dashed lines, while the black solid lines are their combination.

As first done in Hermsen et al. (2016) for the mode-switching pulsar PSR B1822$-$09, we  can easily generalize the 2D-ML method to take into account also the pulse phase information of  the events.
In this 3D-ML approach, we can   sort the events  according to their spatial coordinates and pulse phase  (x,y,$\phi$). The expectation value  of bin (i,j,k) can now be written as \\

 $\mu_{ijk}=\beta+\sigma_u \cdot {PSF}_{ij} + \sigma_p \cdot {PSF}_{ij} \cdot \Phi_k$.

\medskip
\noindent
Here the value of the normalized pulse profile at bin $k$ is represented by $\Phi_k$, while $\sigma_u$ and $\sigma_p$ 
correspond to the unpulsed and pulsed component scale factors, respectively.  
The pulsed fraction $\eta$ can be  determined as $\eta=1/(1+ (N_{\phi} \cdot \sigma_u/\sigma_p))$, with $N_{\phi}$ the number of bins of the normalized pulse profile. 
As shown in  Sect.~\ref{sec_timing}, the X-ray pulse profile of \psr\ is well described by a sinusoid with the same phase in both radio modes. We assume in the following an energy-independent sinusoidal pulse profile.

With this method we derived the pulsed fraction as a function of energy for the Q- and B-mode  shown in Fig.~\ref{fig_pf}.
The Q-mode pulsed fraction steadily increases from $\sim 0.21$ in the  0.2-0.5 keV band to $\sim 0.69$  at 2 keV (consistent with the values reported by \citet{her13}), and then drops 
significantly. This behaviour is different from that shown in the B-mode, in which the pulsed fraction is constant across the 
0.2-2 keV energy band, with an average 0.5-2 keV value of $38\pm5$\% and even a significant measurement of $35\pm11$\% 
in the 0.2-0.5 keV band. The 0.5-2 keV pulsed fraction for the Q-mode is $52\pm4$\%. 

We also performed a spectral analysis of the unpulsed  and pulsed components of  both modes using the source counts per energy bin extracted with  the 3D-ML approach. The results are summarized in  Table~\ref{tab_ML3D}.  
For the Q-mode (10 spectral bins  the 0.2-10 keV band) we found that the pulsed component is properly described by a blackbody (Fig.~\ref{fig_sp}a; red data points; red solid line for model), while a power-law is rejected ($\chi_{\nu}^2/dof=5.09/8$). On the other hand, the unpulsed component (Fig.~\ref{fig_sp}a; blue data points; blue solid line for model)  is well fit by a power-law, while a blackbody is rejected ($\chi_{\nu}^2/dof=3.53/8$). 
The sum of the two fits is indicated by a purple solid line in Fig.~\ref{fig_sp}a, which coincides within uncertainties with the total spectrum derived with the 2D-ML method.

For the B-mode the characterisation of the spectra of the unpulsed- and pulsed components is more difficult due to the lower number of events (by $\sim$ 35\%) of the pulsed and unpulsed emission. Therefore, we used only 7 spectral bins. 
The unpulsed and pulsed components can be fit equally well by either a blackbody  or a power-law  (see the last 4 columns of Table~\ref{tab_ML3D}). 
Fig.~\ref{fig_sp}b  shows the spectra of the pulsed and unpulsed emissions of the B-mode, assuming the same characteristics  found for the Q-mode, i.e. thermal  pulsed emission and non-thermal  unpulsed emission; the red data points and solid line represent the (unabsorbed) pulsed flux measurements and best-fit blackbody model, respectively, and the  blue data points and solid line the unpulsed flux measurements and best-fit power-law model. The purple solid line is  the sum of the two models. The match with the 2D-ML results is less obvious than for the Q-mode case (note that the error regions are substantial),  which is not strange given the degenerate nature of the B-mode spectrum.

\begin{figure}
 \hspace{-1cm}
\includegraphics[angle=-90,width=11cm]{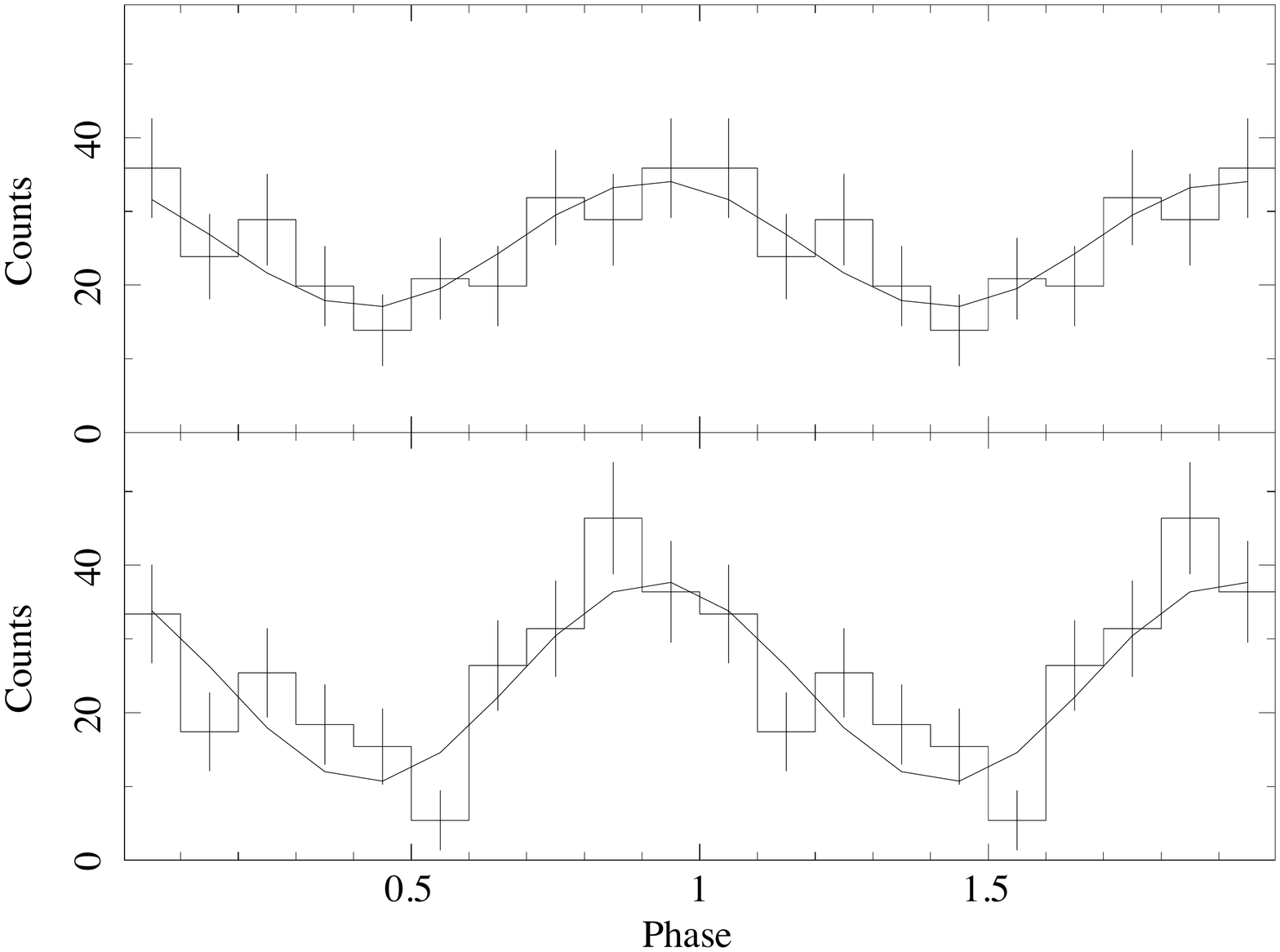}
\caption{\label{fig_lcBBH3}   Folded pulse profiles in the 0.5-2 keV range  for the early-B (top) and late-B (bottom) time intervals (see text). The background has been subtracted. The curves show  fits with a constant plus a sinusoid with phase fixed to that of the whole B-mode. The amplitudes of the sinusoids  are  8.6$\pm$1.8     and  13.7$\pm$1.9 for the early-B and late-B, respectively, and the constants are     25.6$\pm$1.3 and  24.2$\pm$1.4 .}
\vspace{0.5cm}
\end{figure}

\begin{figure}
\hspace{-1cm}
\includegraphics[angle=-90,width=11cm]{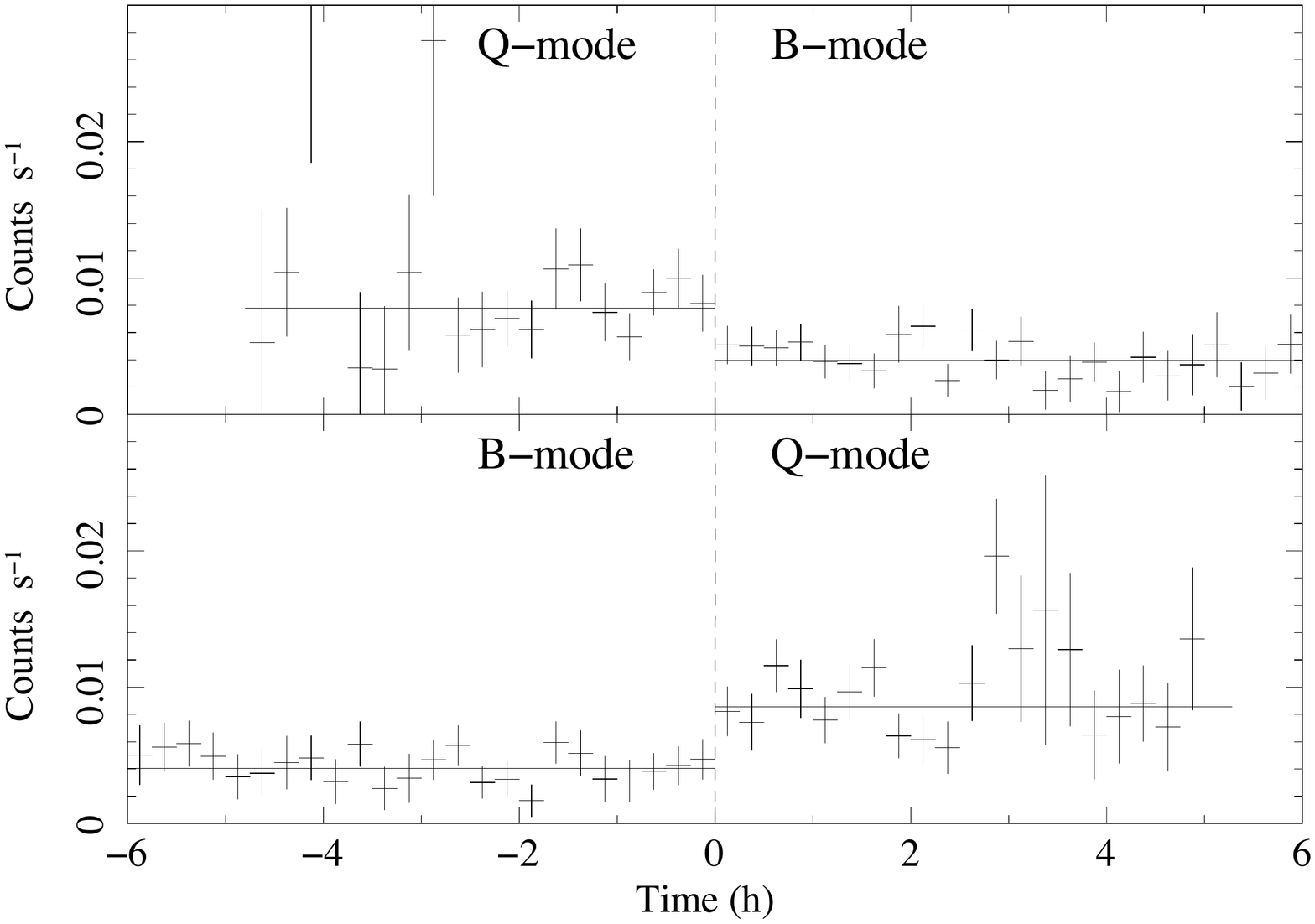}
\caption{\label{fig_transitions}   Light curves  (pn+MOS, background subtracted, 0.2-2.5 keV,  time bins of 900 s) obtained by stacking all the Q to B (top panel) and B to Q (bottom panel) mode transitions  observed in   2014.  The light curves are corrected to take properly into account the exposure time resulting from the different lengths of the  time intervals and from the gaps due to the data cleaning. The horizontal lines indicate fits with constant count rates to 6 hr long intervals and have the following values. 
Q-mode before transition: 0.0078$\pm$0.0006  ($\chi^2_{\nu}$=0.90);
B-mode after transition:    0.0040$\pm$0.0003  ($\chi^2_{\nu}$=0.80);
B-mode before transition: 0.0041$\pm$0.0003  ($\chi^2_{\nu}$=0.66); 
Q-mode after transition:   0.0086$\pm$0.0005  ($\chi^2_{\nu}$=1.17).}
\end{figure}

\subsection{Evolution of X-ray properties   within the radio modes}
\label{sec_evo}

Radio studies showed that, while the Q-mode emission is steady and largely chaotic, the highly-organised B-mode evolves in its profile and subpulse behaviour \citep{ran06}.
The evolution of the radio properties begins immediately after the Q to B transition and is largely accomplished within two hours, 
as can be seen in the radio charts displayed in the Appendix. 
Motivated by these results, we  searched for a possible   evolution of the   X-ray properties.

We divided all the B-mode data in two subsets:  one with all the data collected within 3 hours after the B-mode start (early-B in the following)  and one with the remaining data (late-B). We excluded  the initial 3 hours of the observations that started with the pulsar in B-mode but for which the transition time was not known.  Although the radio evolution occurs on a shorter timescale \citep{bil14}, our  choice of 3 hours was dictated by the need of a comparable counting statistics in the two subsets (351 and 362 pn+MOS counts in the early-B and late-B intervals, respectively). 

Analysing the data as  in Section \ref{sec_timing}, we found that the pulsations in the 0.5-2 keV   range are detected with a significance of only  2.6$\sigma$ in the early-B subset, while their significance in the late-B subset is of 4.6$\sigma$. 
By means of Monte Carlo simulations, assuming the average pulsed fraction of the whole B-mode data derived in Section \ref{sec_timing}, we  found that the   probability   to obtain a significance as small as that observed in   the early-B data is   $\sim$6\%. 
The  pulse profiles of the two subsets, plotted in Fig.~\ref{fig_lcBBH3},  show   a different degree of modulation, also confirmed by  the pulsed fractions obtained with the
3D-ML analysis:  27$\pm$8\% for the early-B and 42$\pm$8\%  for the late-B. 
These results suggest that there might be an evolution of the X-ray properties during the B-mode,  but the statistics are  too low to draw firm conclusions.  A similar analysis to search for variations in the pulsed  fraction during the Q-mode gave negative results.

To search for possible evolution in the X-ray flux,  we also constructed stacked light curves of the mode transitions by summing all the  data. 
These are plotted in Fig.~\ref{fig_transitions}, where we have set the origin of the time axis to the time of the mode transitions.
The count rates before and after the transitions are well fit by constant functions and there is no evidence for any gradual increase/decrease of the flux leading to  or following the mode-switch times. 
The changes in the   X-ray flux occur on a time scale shorter than the bin time of 900 s used in these lightcurves.

\subsection{Search for diffuse X-ray  emission}

To search for the presence of  diffuse emission around \psr\ on angular scales smaller than those resolved with \xmm  , 
we carried out an observation with the {\it Chandra} satellite on 2016 January 15 (Obs. ID 16759). The observation was done with the ACIS-S instrument in Timed Exposure full-imaging mode (frametime: 3.14~s) and lasted 50 ks. The data reduction and analysis were perfomed with the CIAO analysis software v.~4.8 and the point-spread function (PSF) simulation package ChaRT/MARX (v.~5.2), using the calibration files in the CALDB database v.~4.7.1.

\src\ was clearly detected with a net count rate of $(1.0\pm0.1)\times10^{-3}$ cts s$^{-1}$ in the 0.3--8 keV energy range. We extracted a spectrum using the counts in a circle of radius 1.25~arcsec (encircled energy fraction: EEF~$\sim95\%$) for the source and an annulus with radii 5 and 10~arcsec for  the background.   The spectrum is well fit by a blackbody model with absorption and temperature fixed to the values derived with {\it XMM-Newton} for the B-mode spectrum. The resulting flux is $(4.7\pm0.8)\times10^{-15}$ erg cm$^{-2}$ s$^{-1}$  (0.5--2 keV), consistent with that of the B-mode.  A fit with the blackbody plus power-law model with parameters fixed to those of the Q-mode is unacceptable, unless a rescaling of a factor 0.47$\pm$0.07 is applied. Therefore, we conclude  that \psr\ was in the B-mode  during most of (or possibly all)  the {\it  Chandra}  observation.

The radial distribution of the source counts around the pulsar position is consistent with the instrumental PSF, with no evidence for diffuse emission at radii larger than $\sim2''$ (EEF approaching 100\%). We derived 3$\sigma$ upper limits of $3\times10^{-6}$ and $1.5\times10^{-6}$ cts s$^{-1}$ arcsec$^{-2}$ on the average surface brightness within radii of  5$''$ and 10$''$, respectively.
For a power-law spectrum with photon index 2.5, these values correspond to flux upper limits of $1.0\times10^{-15}$  and $2.3\times10^{-15}$ erg cm$^{-2}$ s$^{-1}$  (0.5--2 keV).   The limits derived for larger nebulae are less constraining: for example, in the case of   diffuse emission extending to 20$''$, the  surface brightness upper limit is  $10^{-6}$ cts s$^{-1}$ arcsec$^{-2}$, corresponding to $7\times10^{-15}$ erg cm$^{-2}$ s$^{-1}$.

\begin{figure}
\vspace{-2cm}
\includegraphics[angle=0,width=9cm]{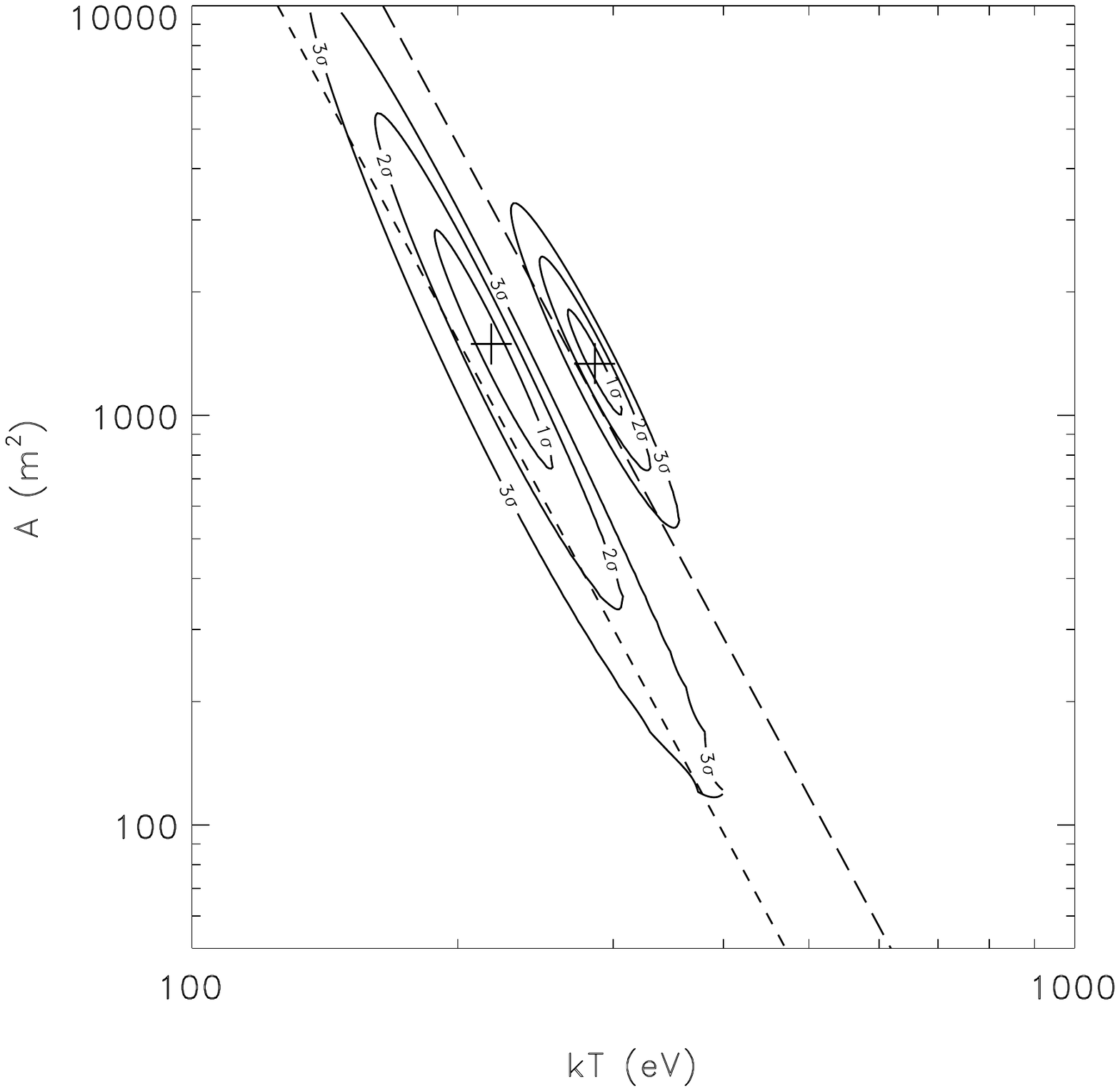}
\vspace{-3cm}
\caption{\label{fig_cont_kt_area}  Confidence regions (1$\sigma$, 2$\sigma$, and 3$\sigma$ c.l.) of blackbody temperature and emitting area for the pulsed thermal component of the B-mode (bottom-left contours) and Q-mode (top-right contours). The dashed lines correspond to bolometric luminosities of 10$^{29}$ and $3\times10^{29}$ erg s$^{-1}$, for a distance of 630 pc.}
\vspace{0.5cm}
\end{figure}

\section{Discussion}

Before discussing a plausible physical interpretation of our results on \psr , 
we  summarize the new  findings of the 2014 campaign  and their robust and model-independent implications. We found that:

\begin{itemize}

\item {\it    X-rays are significantly modulated at the pulsar spin period  also during B-mode}.
Pulsations had been previously   detected only in   Q-mode,
while in   B-mode 3$\sigma$ upper limits on the pulsed fraction   of 56\%  (0.6--1.3 keV) and  46\%   (0.15--12 keV) were set   \citep{mer13}.  
This result is inconsistent with the possibility that   the B-mode   X-ray emission is entirely non-thermal and unpulsed, as it was proposed by \citet{her13}.

\item {\it The energy-dependence of the pulsed fraction differs between the two  modes}.   In   B-mode the pulsed fraction is constant over the 0.2--2 keV range, while in Q-mode it increases with energy, from  a value of $\sim$20\% at the softest energies up to  70$\pm10$\% at 2 keV and then it drops below 50\%.

 \item  {\it  The total X-ray spectrum during B-mode is well described by a single blackbody, while it cannot be fit  by a  single power-law}.  
Even if not required from a statistical point of view (contrary to the Q-mode case), the data are consistent with the presence of an additional component, e.g. another blackbody or, more likely,  a power law, as we discuss below.
 
\item   {\it In Q-mode the pulsed emission is well fit by a blackbody (and not by a power-law), while the unpulsed emission is well fit by a power-law (and not by a blackbody)}.  The best-fit parameters of the pulsed and unpulsed emission are consistent, within the errors, with those of the blackbody and power-law components required to fit the total Q-mode spectrum,  confirming the   results of \citet{her13}.

\item   {\it In B-mode both the pulsed and the unpulsed emission are   well described by  either a  blackbody or a power-law}. Owing to the low statistics, it is impossible to discriminate between these models based only on these spectral fits. However,  the Q-mode results and the physical considerations described below, lead us to favour a scenario in which the B-mode emission consists of pulsed thermal plus unpulsed non-thermal X-rays.
 
\item   {\it There is no evidence for diffuse X-ray emission on angular scales from a few to several arcseconds}.  The upper limit on the diffuse flux within a radius of  5$''$ excludes a major contribution from a pulsar wind nebula to the unpulsed luminosity.

\end{itemize}

These findings allow us to exclude  two simple scenarios that were previously  proposed on the basis of the short 2011 observations, i.e. that  the reduced flux of the B-mode is due to  the disappearance either of  the pulsed thermal  component   \citep{her13} or of a pulsed non-thermal component \citep{mer13} seen  during the Q-mode. 
 
With the caveat that other interpretations cannot be excluded by the current observations,   in the following we concentrate on a plausible scenario in which  thermal and   non-thermal X-rays are emitted in both radio modes of \psr .

 \subsection{A thermal plus non-thermal scenario for the two radio modes}

The simultaneous emission of  thermal and a non-thermal X-rays  is a  common property seen in  all the rotation-powered pulsars that have been studied with sufficient sensitivity.  The thermal emission from young and middle-aged neutron stars results from internal cooling and involves a relatively large fraction of (or the whole) star surface. In older objects, like \psr , only small regions at the magnetic poles are sufficiently heated by backward-accelerated magnetospheric particles to significantly emit in the X-ray band. 

The presence of  thermal and  non-thermal X-rays during Q-mode was already clear in  the 2011 data and is confirmed by the new observations, which yield fully compatible spectral parameters.  Furthermore, the analysis of the Q-mode data with the powerful 3D-ML method shows that the pulsed component is well fit by a blackbody and not by a power-law,  as  found by \citet{her13}. The opposite is true for the  unpulsed component, which is clearly inconsistent with a blackbody. Therefore, we can describe the  Q-mode emission as the sum of two contributions with similar fluxes in the 0.5-2 keV range: a pulsed thermal component, with blackbody temperature kT$\sim$0.28 keV and emission radius $\sim$21 m,  plus an unpulsed non-thermal component, with power-law photon index $\sim$2.5.

Although a single blackbody, with   kT=0.23 keV and  emission radius $\sim$30 m,  gives a fully satisfactory fit of the B-mode spectrum, the presence of an additional power law  is fully compatible with the data.  This  yields  a smaller emitting area for the blackbody, similar to that seen in Q-mode.
The 3D-ML analysis indicates that in B-mode both the pulsed and unpulsed spectra can be fit equally well by either a power-law or a blackbody (see Table~\ref{tab_ML3D}).  However, from a physical point of view, it is quite contrived to  invoke a mechanism able to rapidly switch-off (or strongly suppress) the pulsed thermal component seen in   Q-mode and replace  it with a non-thermal pulsed emission. 
It is  much more natural to explain  the B-mode pulsed flux as  thermal emission closely related to   that seen in the Q-mode, and possibly of the same origin.  This is also supported by the fact that the shape of the  pulse profile does not change, being  nearly sinusoidal  with the maximum very close in phase to the main radio pulse in both modes.

It is interesting to note that the best-fit value of the  temperature  is slightly higher during the Q-mode.  The 30\% change in temperature corresponds, for a fixed emitting area, to an increase of bolometric luminosity of a factor $\sim$3,  consistent with the measured flux increase of 2.9.
This increase of the thermal flux can be compared to the factor $\sim$2 increase in the non-thermal luminosity.
The error regions for the parameters of the blackbody components  (Fig.~\ref{fig_cont_kt_area}) are consistent with either a change in emission area (with fixed temperature) or  a change in temperature (with fixed area),
although the best-fit values favor the latter possibility.

\subsection{Physical interpretation}

Drifting radio subpulses during the B-mode of \psr\ serve as a classical example for associating the phenomenon with the E$\times$B drift of a system of sparks which are generated at the inner accelerating region (IAR). 
The theory was first suggested by \citet{rud75},  who conjectured the IAR to be an inner vacuum gap (IVG).
If  {\bf $\Omega  \cdot B_s$}  $<$0 (where $\Omega = 2 \pi /P$, $P$ being the pulsar period and  $B_s$  the surface magnetic field),  an IVG can be formed just  above the polar cap, where a high potential difference exists. 
Such a region can discharge as a system of sparks by the process of magnetic pair creation. 
The electric field in the IVG  separates   positrons and electrons. The latter accelerate back to the polar cap and heat the surface sufficiently enough to generate soft X-ray thermal emission. 
The positrons instead accelerate away from the stellar surface and produce secondary pairs, which eventually leads to  non-stationary spark-associated plasma columns, where coherent radio emission is generated at a distance of about 50 stellar radii. 
The  plasma columns undergo a E$\times$B drift in the IAR, which is reflected in the radio emission region.

\citet{des01} were able to model the drifting radio features observed in the B-mode of \psr\ as a system of 20 sparks rotating in the outer edge of the pulsar beam as a carousel, with a circulation time of 37$P$.  
The fact that this circulation time was slower than that predicted by \citet{rud75}, led to the refinement of the IAR model as a partially
screened   gap   \citep{gil03}.  
The basic feature of the  partially screened   gap (PSG) model is that the polar cap is maintained just above a critical temperature $T_c$ where positive ions can be extracted from the surface, which can then screen the vacuum electric field and reduce the drift speed.

Our blackbody fit to the pulsed B-mode spectrum gives an emitting area $A_s$ of  $\sim$1500 m$^2$,  two orders of magnitude  smaller than that expected for the polar cap in a  dipole geometry  ($A\sim \pi R_{pc}^2 = \frac {2 \pi^2 R^3}{c P} = 1.3\times10^5$ m$^2$, for a star radius R=13 km). This suggests that  only a small fraction of the polar cap is heated or that the magnetic field is non-dipolar at the
polar cap surface. 
Indeed such non-dipolar fields are  expected in pulsars, 
as it has been shown by \citet{gil01,gil02a,gil02b}.
If the field near the surface is not dipolar,  the ratio of the surface to dipolar field would be $B_s/B_d \sim100$, by flux conservation,  implying $B_s \sim4\times10^{14}$ G.  \citet{gep13} showed that Hall drift of the crustal magnetic field can generate such strong fields.

No  clear drifting radio features are seen during the Q-mode, so no further analysis regarding the radio and X-ray connection can be established, although, once again, the X-ray data appear to strongly suggest a very small blackbody emitting area, similar to that of the B-mode.

The next important phenomenon is the  change of the pulsar radio emission from B to Q-mode, where the tip-over occurs over perhaps hardly more than a single  period. 
In the context of inner gap models, \citet{zha97} proposed that the mode switching is caused by transitions between two states  in which  inverse Compton scattering plays a major role in causing the inner gap breakdown. At the mode-switch, the gap height and Lorentz factor of the primary particles change rapidly.  The transitions can occur when the pulsar surface temperature is close to a critical value T$_c$, which,     for    \psr\ and in the case of  a multipolar magnetic field, is predicted by these authors to be T$_c$=10$^6$.  Considering the uncertainties involved both on the theoretical and observational side, this value is in good agreement with our results.

More recently, \citet{sza15} suggested that the effect  could arise due to two states of the PSG, termed as PSG-off and PSG-on state.
In the PSG-off state the IAR is initially vacuum and pair cascades due to curvature radiation  can operate. 
The backflowing particles then hit the surface and raise the temperature to $T_c$, where the ions  are extracted. 
In the PSG-on state the surface is already at a temperature slightly above $T_c$ and the dominant process of pair cascade is due to inverse Compton scattering. 
The ions which are now extracted from the star can completely screen the electric field and no further heating can take place. 
Thus the prediction is that in both PSG-off and PSG-on state the temperature of the polar cap is $\sim T_c$. 
One also does not expect a change in the surface area of the polar cap between the PSG on/off  states. 
The change in the intensity of the X-ray blackbody  appears to be rather perplexing in this regard, thus ruling out a  simple application of the PSG on/off model. 

The presence of the Q-mode precursor  in the radio might    be  invoked to find clues about some additional source of change in area or temperature needed to account for  the higher blackbody flux. 
If the precursor emission is associated with the closed field lines, that could lead to an increased area,
while if the precursor emission is associated with an  increased return current, that causes a temperature increase.

It has been argued that mode changes can be seen as changes in state of the global magnetospheric configuration. Magnetosphere models can be built which are physically self-consistent solutions of the highly non-linear equations \citep{goo04,tim06}, including a range of configurations for the last closed fieldline and hence possible polar cap boundaries. However, these are steady-state models and give no hint as to what triggers a change from one state to another. Nor do they give the timescale of the mode durations and - especially - explain why certain states are preferred over others (and why, it seems, only two?).

Radio observations of other pulsars give clues that mode-changing involves not only the polar cap but the whole magnetosphere. Notably, \citet{kra06},  \citet{lyn10}, \citet{cam12} and \citet{lor12b} 
reported evidence for a relation between radio mode switching and changes in period derivative suggesting abrupt changes in the rate of angular momentum loss caused by sudden changes in magnetospheric currents along the open field lines. Such changes in magnetospheric currents could presumably be responsible for switches in non-thermal magnetospheric emission as well as in emitted thermal X-rays due to enhanced heating or changes in hot spot area.
Furthermore, PSR B1822--09,   a nearly orthogonal rotator in which we 
can observe both poles, 
has two distinct modes: one with a regular pulse modulation and one largely chaotic in a manner resembling \psr\ \citep{gil94,lat12}.
The modulations are in phase at both poles and the modes change rapidly and near-simultaneously at both poles on a timescale of minutes. This suggests inter-pole communication. Similar phase-locked pulse modulation between poles has also been found in PSR B1702--19  and PSR B1055--52 \citep{wel07,wel12}.

It may be that communication between poles is only possible when a pulsar is highly inclined, but it may simply be that this effect can only be observed when both poles are visible. If communication is possible for any modulating or mode-changing pulsar, we may speculate that it is also occurring in \psr . The obvious line of communication would be via the magnetic fieldlines at or close to the polar cap boundary.  If modulation with the observed precision of the B-mode were occurring at both poles it would challenge the view that the drifting subpulses are the product of  E$\times$B drift in a local multipole configuration (it being already hard to understand how sparks could form a perfect carousel in such an environment) and would require the presence of magnetosphere-wide drift similar to auroral phenomena \citep[e.g.,][]{wri03}.

If the E$\times$B drift is a global phenomenon, one might expect a precise short $\approx2P$ modulation, such as that found in \psr , only to be sustainable for any length of time in the near-symmetry of a nearly-aligned pulsar magnetosphere. Thus, although the B-mode can be seen as a steady magnetospheric eigenstate, the specific fieldlines of communication between the poles would themselves gradually evolve as a result of the slight misalignment of the rotation and magnetic axes. Eventually the delicately-maintained eigenstate would suddenly collapse and a chaotic Q-mode ensue until the B-mode eigenstate could assert itself again. In this picture the observed temperature difference between the modes might be a consequence of the mode change and its evolution, rather than its cause, and the same could be argued for the non-thermal emission.

 \subsection{Modelling of the pulsed fraction}
 \label{sec_pf}
 
The 3D-ML analysis has shown that the pulsed component in the Q-mode is thermal, 
and the same conclusion may also hold for the B-mode, although with less certainty. 
The high pulsed fraction of this component is rather puzzling, since the radio emission properties have long suggested that \psr\ is an aligned rotator, and hence the polar cap hot spot should not give rise to  strongly pulsed X-rays.

To  address this issue in some quantitative detail we computed the radiation intensity $I(E,\,\mu)$ emerging from a
plane-parallel, magnetized atmosphere with $\mathbf B$ along the vertical axis, as appropriate if the hot cap is at the  magnetic pole;
here $E$ is the photon energy at the surface of the star and $\mu$ is the cosine of the angle between the surface normal and the ray
direction. The magnetic field strength is $B=4\times 10^{12}\ \mathrm G$ and we take an effective temperature, at infinity, of
$T_\mathrm{eff}=250\ \mathrm{eV}$;  
the star mass and radius are $M=1.4\ M_\odot$ and $R=13\ \mathrm{km}$, respectively. 
The atmosphere composition is pure H and complete ionization is assumed (see \citet{zan06}, and references therein). $I(E,\mu)$ is then used to derive the pulse profiles, by mean of the technique described in \citet{tur13}, which accounts for general-relativistic ray-bending and applies to circular spots of arbitrary size. Actually, since here the cap radius is $\ll R$, 
the emitting spot is virtually point-like and the use of a single atmospheric model is fully justified.
In order to make the comparison with observational data meaningful, in computing the pulsed fraction the contribution of an unpulsed power-law component with the  appropriate spectral parameters was taken into account. 

The light blue region in Fig.~\ref{fig_pf} shows the computed values of the pulsed fraction 
for the most likely ranges of the two geometrical angles, $5^\circ < \chi < 10^\circ$ and $10^\circ < \xi < 15^\circ$ (\citet{des01}, $\xi$ is the angle between the magnetic and rotation axis, $\chi$ is the angle between the rotation axis and the line of sight).
The computed pulsed fraction increases with energy then drops above $\sim 2$ keV where the power-law flux becomes dominant, but 
it is systematically too low to account for the observed values, especially in the Q mode.
In a similar computation, \citet{sto14} adopted an atmosphere model with 
$T_\mathrm{eff}\sim 150\ \mathrm{eV}$. We recomputed the pulsed fraction following the same 
approach outlined above, but with an effective temperature of $140\ \mathrm{eV}$. 
While the pulsed fractions in this case are somehow higher (darker region in Fig.~\ref{fig_pf}), they still fail to account for the observed ones.

The pulsed fractions computed by Storch et al., corrected for the presence of an unpulsed component,
are closer to the data, although they still underpredict the observed values (see the dashed line in Fig.~\ref{fig_pf}). The main difference
between the two approaches is in our assumption of complete ionization, while \citet{sto14} used a partially ionized atmosphere. 
Surprisingly, the effect on the pulsed fraction is quite substantial, despite the fraction of neutral H atoms for the
typical values of the temperature and magnetic field relevant to \psr\ is well below $0.1\%$ \citep{pot04}.

Given the uncertainties in  parameters ($T_{\mathrm eff}$, $B$) and in the physical conditions of the atmosphere, one may tentatively
conclude that magnetic beaming in an atmosphere on top of the heated polar cap can reproduce the large observed pulsed fraction,
and its dependence on   energy, for values of the two geometrical angles $\xi$ and $\chi$ in the range derived by radio
observations.

\subsection{Possible evolution during B-mode}

Finally, having achieved a reasonable characterisation of the X-ray components in the B- and Q-modes, we turned our attention to see if our extensive observations could tell us anything about the evolution of these characteristics {\it within} the modes. 
A noticeable change in the pulsar radio profile shape and in the drifting feature occurs during the course of the B-mode \citep{bac11,bil14}.
The work by \citet{bac11}  revealed that, based on the carousel model, the drift change could be produced by a 10\% increase in the average number of sub-beams and a 16\%   increase in the carousel circulation time.  
They speculated that under the PSG model the increase in circulation time should be related to a temperature change of about 1.4\% across the B-mode.  
Such a small change is  clearly  consistent with our X-ray data, but unfortunately cannot be detected with the current instrumentation.

In section \ref{sec_evo} we presented marginal evidence for an increase of the pulsed fraction during B-mode.
We note that, if  this occurs on the same timescale of the evolution of  the radio properties, our analysis underestimates its significance 
since the optimal boundary, with more plentiful X-ray counts, would have been, say, 1.5 hours, rather than the selected 3 hours. 
The failure to detect pulsations in B-mode during the 2011 observations, in  which the later stages of the  B-mode were poorly covered \citep{her13},  might be taken as an other indication in favor of an evolution of the pulsed flux.
If this effect is confirmed with more X-ray data, its study can provide interesting clues for the  understanding of the  processes responsible for the mode-switching behavior of \psr .

\section{Conclusions}

Thanks to the long duration of    the  \xmm\ Large Program  with simultaneous LOFAR, LWA and Arecibo radio monitoring carried out in 2014, we  obtained several new  results on the  X-ray emission  of the prototypal bimodal radio pulsar \src . 
Though the cause for the X-ray variability correlated with the radio modes remains unknown, we could explore in much more detail some of the scenarios that were proposed  to explain the remarkable X-ray spectral and timing properties of this pulsar. In particular, the discovery of pulsations in the B-mode and the failure of a single power-law to fit the B-mode spectrum rule out simple interpretations which tried to explain the difference between the two modes as a change in a single pulsed component, present only during the Q-mode \citep{her13,mer13}. 

We showed that the situation is indeed more complex, and propose a consistent picture in which pulsed thermal and unpulsed non-thermal emission are simultaneously  present in both radio modes and vary in a correlated way.
Such a correlation is not surprising in the framework of the Vacuum Gap models discussed above, since the pairs  are produced relatively close to the star surface and the accelerated particles are responsible both for the non-thermal emission in the magnetosphere and for the heating of the polar regions through return currents.
A quantitative assessment of the relation between thermal and non-thermal emission by comparing different neutron stars is complicated by the presence of other factors, e.g. orientation, magnetic field strength, age, which introduce a variance in the observed properties.  In the case of \psr\ we have instead the unique possibility to examine the relation between these components without such complications.

On the other hand,  Space Charge Limited Flow models with particles leaving freely the star surface  \citep[e.g.,][]{aro79,zha00}, predict
the pair-creation front much higher in the magnetosphere. Since in this case only a few percent of the return particles might reach the surface, one would not naturally expect a correlation beteween the thermal and non-thermal emission.
   
Finally, we found some evidence for an evolution of the timing properties during the B-mode, which,
if confirmed, can provide another important diagnostic to study the correlation between radio and X-ray properties in \psr\ and   shed light on the physical processes responsible for the mode switching behavior. \\

\acknowledgements

We thank Norbert Schartel and the staff of the  {\it XMM-Newton} Science Operation Center, in particular Jan-Uwe Ness,  for their great support in the scheduling of  these time-constrained observations. We acknowledge Aris Karastergiou for his contributions to the FR606 observing system.

{\it XMM-Newton} is an ESA science mission with instruments and contributions directly funded by ESA Member States and the USA. 

LOFAR, the Low-Frequency Array designed and constructed by   ASTRON, has facilities in several countries, that are owned by  various parties (each with their own funding sources), and that are collectively operated by the International LOFAR Telescope (ILT)  foundation under a joint scientific policy.
Part of this work is based on observations with LOFAR telescopes of the German Long-Wavelength consortium. Specifically, we acknowledge use of the data recording hardware provided by the Max-Planck-Institut f{\"u}r Radioastronomie in Bonn (MPIfR) and the use of the international LOFAR stations operated by the MPIfR (DE601), the Max-Planck-Institut f{\"u}r Astrophysik in Garching (DE602), the Th{\"u}ringer Landessternwarte in Tautenburg (DE603) and DE605, which is jointly operated by the Ruhr-Universit{\"a}t Bochum and the Forschungszentrum J{\"u}lich. These international LOFAR stations are funded by the Max-Planck-Gesellschaft, the Bundesministerium
f{\"u}r Bildung und Forschung (BMBF) and the German states of Th{\"u}ringen and Nordrhein-Westfalen.
Nan\c{c}ay Radio Observatory is operated by Paris Observatory, associated with the French Centre
National de la Recherche Scientifique and  Universit\'{e} d'Orl\'{e}ans.

Construction of the LWA has been supported by the Office of  Naval Research under Contract N00014-07-C-0147.  Support for operations and continuing development of the LWA1 is provided by the   National Science Foundation under grants AST-1139963 and AST-1139974  of the University Radio Observatory program.

The Arecibo Observatory is operated by SRI International under  a cooperative agreement with the National Science Foundation  (AST-1100968), and in alliance with Ana G. Mendez-Universidad   Metropolitana, and the Universities Space Research Association.

The financial assistance of the South African SKA Project (SKA SA) towards this research is hereby acknowledged. Opinions expressed and conclusions arrived at are those of the authors and are not necessarily to be attributed to the SKA SA.

SM, AT, PE, RT and AP acknowledge financial contribution from  PRIN INAF 2014.

JWTH acknowledges funding from an NWO Vidi fellowship and from  the European Research Council under the European Union's Seventh   Framework Programme (FP/2007-2013) / ERC Starting Grant agreement   nr. 337062 ("DRAGNET").

JMR acknowledges funding from US NSF grant 09-68296 and a NASA Space Grant.

PE acknowledges funding in the framework of the NWO Vidi award A.2320.0076 (PI: N. Rea).

SO is supported by the Alexander von Humboldt Foundation. \\

Facilities: \xmm , $Chandra$, LOFAR, LWA, Arecibo

\bibliographystyle{aa}

\section{Appendix}

\begin{figure*}
\includegraphics[angle=0,width=17cm]{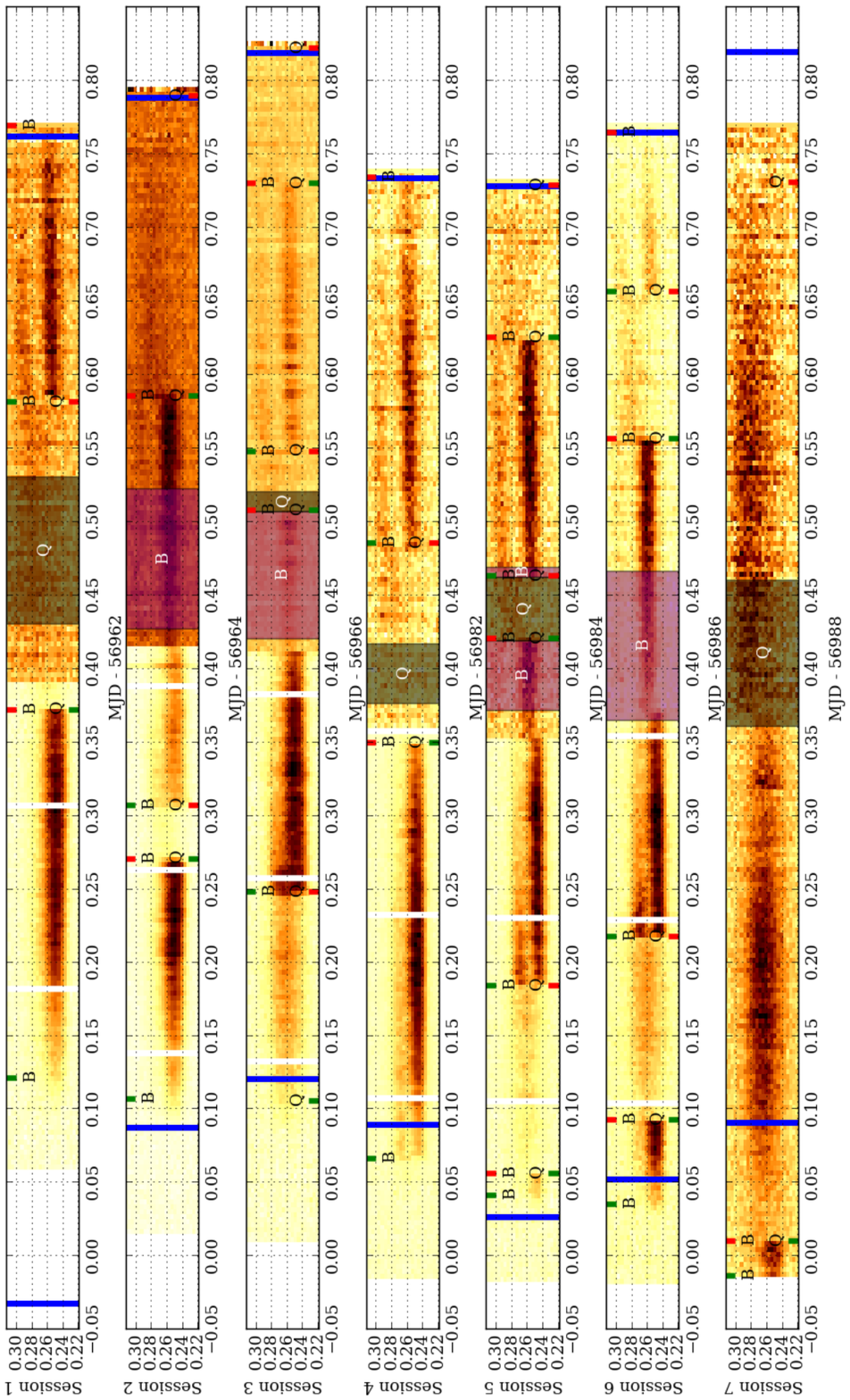}
\vspace{-1.5cm}
\caption{\label{fig_app} Summary of the LOFAR, LWA and Arecibo radio data of the seven
  {\it XMM-Newton} observing sessions.   }

\end{figure*}

In Fig.~\ref{fig_app} we present a summary of the LOFAR, LWA and Arecibo radio data of the seven
  {\it XMM-Newton} observing sessions.
 In each sub-panel, the radio
  pulsed intensity is shown as a function of rotational phase (only
  0.1 in rotational phase around the main pulse peak) and time.  The
  vertical blue bars indicate the start/stop times of the {\it
  XMM-Newton} observations.  Green/red ticks denote the start/stop
  times of mode instances.  B/Q-mode ticks are at the
  top/bottom of the sub-panels.  Each session begins with LOFAR
  observations at $\sim 150$\,MHz, followed by LWA observations at
  $\sim 60$\,MHz.  Note that the pulse profile shape and noise
  properties of these data sets are quite different.  The $\sim
  350$-MHz Arecibo data are not shown directly, but shaded areas
  denoted `B' and `Q' indicate the range of times in which the Arecibo
  data show PSR B0943+10 to be in one of these two modes.  The mode
  determinations from the overlapping Arecibo data agree with what is
  inferred from LOFAR and LWA.
  
\end{document}